\documentclass[authoryear,final,review,12pt]{elsarticle}
\usepackage[lmargin=1in, rmargin=1in, tmargin=1in, bmargin=1in, nohead]{geometry}

\usepackage{multirow}

\usepackage{amssymb,amsmath}
\usepackage{amsthm}
\usepackage{rotating}

\usepackage{epsfig}
\usepackage{epsf}
\usepackage{psfrag}
\usepackage{caption}
\usepackage{latexsym, graphics, psfrag, amscd, amssymb, pb-diagram}
\usepackage[bottom]{footmisc}
\biboptions{authoryear}

\usepackage{color}
\usepackage[all,cmtip]{xy}

\usepackage[lined, ruled, linesnumbered]{algorithm2e}
\usepackage{changepage}
\usepackage{xassoccnt}
\usepackage{lastpage}
\usepackage{hyperref}
\usepackage{float}
\usepackage{setspace}

\newcommand{\myvec}[1]%
{\stackrel{\raisebox{-2pt}[0pt][0pt]{\small$\rightharpoonup$}}{#1}}

\newcommand{\st} {\mathrm{subject~to}}

\NewTotalDocumentCounter{totalfigures}
\NewTotalDocumentCounter{totaltables}
\NewTotalDocumentCounter{appendixchapters}
\DeclareAssociatedCounters{figure}{totalfigures}
\DeclareAssociatedCounters{table}{totaltables}
\newcommand{\totalpages}{\zref@extractdefault{LastPage}{page}{0}}

\usepackage{lipsum}
\makeatletter
\def\ps@pprintTitle{%
 \let\@oddhead\@empty
 \let\@evenhead\@empty
 \def\@oddfoot{}%
 \let\@evenfoot\@oddfoot}
\makeatother

\begin{document}
\singlespacing

\begin{frontmatter}

\title{Predicting Future Cognitive Decline with Hyperbolic Stochastic Coding}

\author[asu]{Jie Zhang}
\author[asu]{Qunxi Dong}
\author[asu]{Jie Shi}
\author[asu]{Qingyang Li}
\author[mc]{Cynthia M. Stonnington} 
\author[iit]{\\Boris A. Gutman}
\author[bai]{Kewei Chen}
\author[bai]{Eric M. Reiman}
\author[mayo]{Richard J. Caselli}
\author[usc]{\\Paul M. Thompson}
\author[um]{Jieping Ye}
\author[asu]{Yalin Wang}
\author[]{and for the Alzheimer's Disease Neuroimaging Initiative*}

\address[asu] {School of Computing, Informatics, and Decision Systems Engineering,\\ Arizona State University, Tempe, AZ, USA}
\address[mc] {Department of Psychiatry and Psychology,\\ Mayo Clinic Arizona, Scottsdale, AZ, USA}
\address[iit] {Armour College of Engineering, \\Illinois Institute of Technology, Chicago, IL, USA}
\address[bai] {Banner Alzheimer's Institute, Phoenix, AZ, USA}
\address[mayo]{Department of Neurology, Mayo Clinic Arizona, Scottsdale, AZ, USA}
\address[usc] {Imaging Genetics Center, Institute for Neuroimaging and Informatics,\\ University of Southern California, Los Angeles, CA, USA}
\address[um] {Department of Computational Medicine and Bioinformatics \& \\Department of Electrical Engineering and Computer Science,\\ University of Michigan, Ann Arbor, MI, USA}

\begin{abstract}
\let\thefootnote\relax\footnotetext{Data used in preparation of this article were obtained from the Alzheimer\textquoteright s Disease Neuroimaging Initiative (ADNI) database (adni.loni.usc.edu). As such, the investigators within the ADNI contributed to the design and implementation of ADNI and/or provided data but did not participate in analysis or writing of this report. A complete listing of ADNI investigators can be found at: https://adni.loni.usc.edu/wp-content/uploads/how\textunderscore to\textunderscore apply/ADNI\textunderscore Acknowledgement\textunderscore List.pdf \\
\textbf{E-mail:} ylwang@asu.edu}\noindent
Hyperbolic geometry has been successfully applied in modeling brain cortical and subcortical surfaces with general topological structures. However, such approaches, similar to other surface-based brain morphology analysis methods, usually generate high dimensional features. It limits their statistical power in cognitive decline prediction research, especially in datasets with limited subject numbers. To address the above limitation, we propose a novel framework termed as hyperbolic stochastic coding (HSC). We first compute diffeomorphic maps between general topological surfaces by mapping them to a canonical hyperbolic parameter space with consistent boundary conditions and extracts critical shape features. Secondly, in the hyperbolic parameter space, we introduce a farthest point sampling with breadth-first search method to obtain ring-shaped patches. Thirdly, stochastic coordinate coding and max-pooling algorithms are adopted for feature dimension reduction. We further validate the proposed system by comparing its classification accuracy with some other methods on two brain imaging datasets for Alzheimer's disease (AD) progression studies. Our preliminary experimental results show that our algorithm 
achieves superior results on various classification tasks. Our work may enrich surface-based brain imaging research tools and potentially result in a diagnostic and prognostic indicator to be useful in individualized treatment strategies.
\end{abstract}


\end{frontmatter}

\section{Introduction}
Alzheimer's Disease (AD), an irreversible neurodegenerative disaese, is the most common cause of dementia among older adults. It is generally agreed that accurate presymptomatic diagnosis and preventive treatment of AD could have enormous public health benefits. Brain structural magnetic resonance imaging (sMRI) analysis has the potential to provide valid diagnostic biomarkers of the preclinical stage as well as symptomatic AD~\citep{pmid20139996}. For example, a single-valued sMRI-based atrophy is used as a neurodegeneration marker in the recently proposed AD descriptive “A/T/N” (amyloid, tau, neurodegeneration) system~\citep{jack2016t} to clinically define AD. Tosun et al. proposed MRI-based approaches to impute Abeta status~\citep{tosun2014multimodal,tosun2016amyloid}. Their results demonstrated that sMRI can be used to predict the amyloid status of MCI individuals and mild AD patients. Recently, brain morphology measures have been integrated with machine learning algorithms to classify individual subjects into different diagnostic groups~\citep[e.g.][]{sun2009elucidating,pmid18228600,wang2013applying,li2014abnormal}. It  offers a promising approach to computer-aided cognitive decline prediction by leveraging both sensitive brain image features and powerful machine learning techniques.

Although most brain sMRI analysis approaches use cortical and subcortical volumes \citep[e.g.][]{pmid12552040,pmid18054253,pmid20375138,pmid20382238}, recent research has demonstrated that surface-based analyses, \citep[e.g.][]{pmid15772166,thompson2004mapping,pmid18228600,pmid20129863,pmid21272654} can offer advantages over volume measures, due to their sub-voxel accuracy and the capability of detecting subtle subregional changes. In surface-based brain imaging research, a practical approach to model brain landmark curves is to model them as surface boundaries by cutting open cortical surfaces along these landmarks. Thus they are modeled as open boundaries to be matched across subjects \citep{Tsui:IPMI13,shi2019hyperbolic} or be used as shape indices~\citep{Zeng2013,shi2017conformal}. Similarly, adding open boundaries have been proved to be useful in modeling ventricular surfaces which have a concave shape and complex branching topology~\citep{Wang:NIMG10,shi2015studying}. We call these genus-zero surfaces with more than two open boundaries as \emph{general topological surfaces} and hyperbolic geometry has been demonstrated to be useful to model general topological surfaces. However, most of current hyperbolic space-based brain imaging methods have been focused on studying group difference between diagnostic groups. To develop brain imaging methods for personal medicine research, it would be advantageous to design powerful machine learning methods that work on general topological surface features for the identification of AD symptoms on an individual basis.

One of the major challenges to directly apply vertex-wise surface features, such as surface tensor-based morphometry (TBM)~\citep{thompson2000growth,chung2008tensor}, to cognitive decline prediction research is that the surface feature dimension is usually much larger than the number of subjects, the so-called \emph{high dimension-small sample problem}. 
Existing feature dimension reduction approaches include feature selection~\citep{fan:miccai05,jain1997feature}, feature extraction~\citep{saadi2007optimally,guyon2008feature,scholkopft1999fisher,jolliffe2011principal} and sparse coding-based methods~\citep{vounou2010discovering,donoho2006,wang2013applying}. In most cases, information is lost when mapping high-dimenstional features into a lower-dimensional space. However, by defining a better lower-dimensional subspace, sparse coding~\citep{lee2006efficient,mairal2009online} may limit such information loss. Sparse coding  has been previously proposed to learn an over-complete set of basis vectors (dictionary) to represent input vectors efficiently and concisely~\citep{pmid16576749}. It has shown to be efficient for many tasks such as image deblurring~\citep{yin2008}, super-resolution~\citep{yang2010image}, classification~\citep{mairal2009online}, functional connectivity~\citep{pmid29993466,lv2015holistic,Lvtask,pmid26466353,LvSparse} and structural morphometry analysis~\citep{zhang2017multi,li2017transcriptome}. However, solving sparse coding remains a
computationally challenging problem, especially when dealing with large-scale datasets and learning large size dictionaries~\citep{lin2014stochastic}.

To generalize sparse coding to process general topological surface features~\citep{mairal2009online}, 
we propose a novel pipeline to extract sparse hyperbolic features for classification termed hyperbolic stochastic coding (HSC), consisting of our unique farthest point sampling with breadth-first search (FPSBS) method for ring-shaped surface patches extraction,  stochastic coordinate coding (SCC) and max-pooling methods for feature dimension reduction, to extract critical low-dimensional shape features from the hyperbolic TBM maps. We call such features as HSC measures. Then the AdaBoost classifier~\citep{freund1997decision} is further adopted on these HSC measures for AD clinical group classification and cognitive decline prediction. We hypothesize that our HSC measures may outperform volume, area and shape-based cortical structural measures on discriminating clinical groups related with AD~\citep{li2014abnormal,pmid18228600,jack1999prediction,leung2010automated} 
We validate our system in a publicly available brain image dataset, Alzheimer’s Disease Neuroimaging Initiative (ADNI) cohort~\citep{ADNI}. With the sMRI baseline data of  133 mild cognitive impairment (MCI) subjects, consisting of 71 MCI converter (MCIc) and 62 MCI stable (MCIs) subjects~\citep{shi2015studying} and 115 subjects (30 AD, 44 MCI and 40 cognitively unimpaired (CU) subjects)~\citep{shi2019hyperbolic}, we set out to test our hypothesis by performing 
classification accuracy comparison with three other popular structural measures (volume, area, and shape-based biomarkers). 

\section{Subjects and Methods}
\subsection{Subjects}
Data for testing the performances of our proposed HSC are obtained from the ADNI database (adni.loni.usc.edu). The ADNI was launched in 2003 as a public-private partnership, led by Principal Investigator Michael W. Weiner, MD. The primary goal of ADNI is to test whether biological markers such as serial MRI and positron emission tomography (PET), combined with clinical and neuropsychological assessments can measure the progression of MCI and early AD. Determination of sensitive biomarkers aids researchers and clinicians to develop new treatments and monitor their clinical effectiveness, as well as lessen the time and cost of clinical trials. The initial ADNI (ADNI-1) database recruited 800 subjects from over 50 sites across the U.S. and Canada and it has been followed by ADNI-GO and ADNI-2. To date, these three databases have recruited over 1500 adults, ages 55 to 90, consisting of elderly cognitive unimpaired individuals, people with early or late MCI, and people with early AD. The follow up duration of each subject is specified in their corresponding  protocols for ADNI-1, ADNI-2 and ADNI-GO. Subjects of ADNI-1 and ADNI-GO had the option to be followed in ADNI-2. For up-to-date information, see www.adni-info.org.

We use two ADNI datasets to validate our system. They are the same datasets used in our prior work~\citep{shi2015studying,shi2019hyperbolic} which mainly studied group differences between different clinical groups. As a generalization of our prior work, the current work studies personalized diagnosis with the same datasets. Studies indicate that ventricular enlargement is an important measure related with AD progression~\citep{shi2015studying,pmid15275931}. In Dataset I, we select 133 subjects from the MCI group in the ADNI-1~\citep{ADNI} baseline dataset as~\citep{shi2015studying,zhang2016hyperbolic}. All subjects have both sMRI and fluorodeoxyglucose positron emission tomography (FDG-PET) data. They include 71 subjects (age: $74.77 \pm 6.81$) who develop incident AD during the subsequent 36 months, which we call the MCI converter (MCIc) group, and 62 subjects (age: $75.42 \pm 7.83$ years) who do not during the same period, which we call the MCI stable (MCIs) group. These subjects are chosen on the basis of having at least 36 months of longitudinal data. If a subject developed incident AD more than 36 months after baseline, it is assigned to the MCIs group. All subjects undergo thorough clinical and cognitive assessments at the time of acquisition, including the Mini-Mental State Examination (MMSE) score, Alzheimer’s disease assessment scale – Cognitive (ADAS-COG)~\citep{rosen1984new} and Auditory Verbal Learning Test (AVLT)~\citep{Rey:1964}. The demographic statistical information of this dataset is shown in Table~\ref{tab:dataset1}. 
\begin{table}[b]
\centering
\begin{tabular}{ccccc} \\
  \hline
Group &Gender (F/M) & Education & Age & MMSE\\\hline
MCIc&26/45&15.99$\pm$2.73&74.77$\pm$6.81&26.83$\pm$1.60\\
MCIs&18/44&15.87$\pm$2.76&75.42$\pm$7.83&27.66$\pm$1.57\\
  \hline
\end{tabular}
\caption{Demographic statistic information of Dataset I.}
\label{tab:dataset1}
\end{table}

In Dataset II, we study cortical morphometry for tracking AD progression. Dataset II has 115 T1-weighted MRIs from the ADNI-1~\citep{ADNI} baseline dataset, including 30 AD patients, 45 MCI subjects and 40 CU subjects~\citep{shi2019hyperbolic}. All subjects underwent through MMSE~\citep{folstein1975mini}. The demographic statistics with matched gender, education, age and MMSE are shown in Table~\ref{tab:dataset2}. 

\begin{table}[t]
\centering
\begin{tabular}{ccccc} \\
  \hline
Group&Gender (F/M) & Education & Age & MMSE\\\hline
AD&15/15&15.22$\pm$2.61&76.22$\pm$7.34&23.07$\pm$2.02\\
MCI&19/26&16.11$\pm$2.56&73.86$\pm$8.20&26.95$\pm$1.34\\
CU &18/22&17.25$\pm$1.90&76.53$\pm$6.02&29.11$\pm$1.03\\
  \hline
\end{tabular}
\caption{Demographic statistical information of Dataset II.}
\label{tab:dataset2}
\end{table}

\begin{figure}[!ht]
\centering
\includegraphics[height=0.7\textheight]{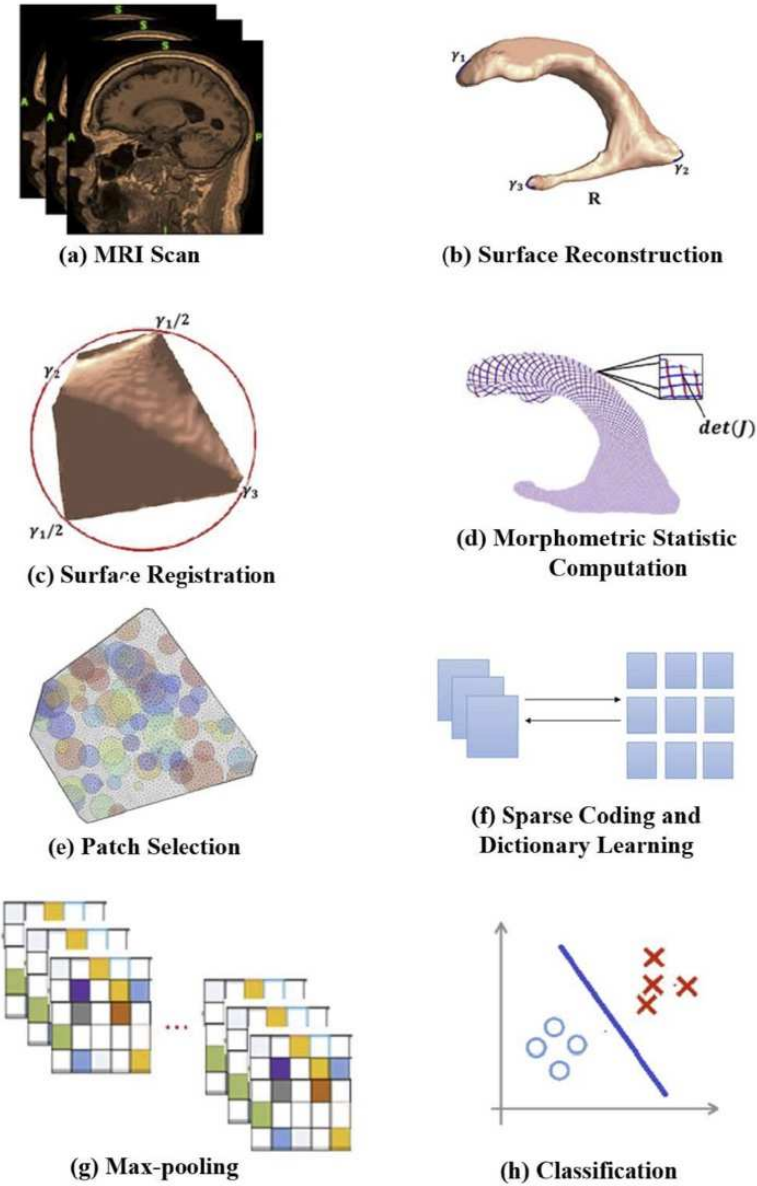}
\caption{The major processing steps in the proposed framework.}
\label{fig:pipeline}
\end{figure}

\subsection{Overview of the Proposed Framework}
The major computational steps of our proposed work are illustrated in Fig.~\ref{fig:pipeline} where we take a left ventricular surface as an example. There are two major stages in the process. In the first stage, we perform ventricular surface reconstruction from MRI data, surface registration and surface TBM feature computation. The second stage is for HSC measure computation. Specifically, we build ring-shaped patches on the hyperbolic parameter space by FPSBS to initialize the original dictionary. Dictionary learning and max-pooling are performed for feature dimension reduction. Following that, Adaboost is adopted to diagnose different clinical groups and predict future AD conversions.

\subsection{Brain Surface Registration with Hyperbolic Ricci Flow and Harmonic Map}
\label{sec:HRF}
\begin{figure}[t]
\centering
\includegraphics[width=5in]{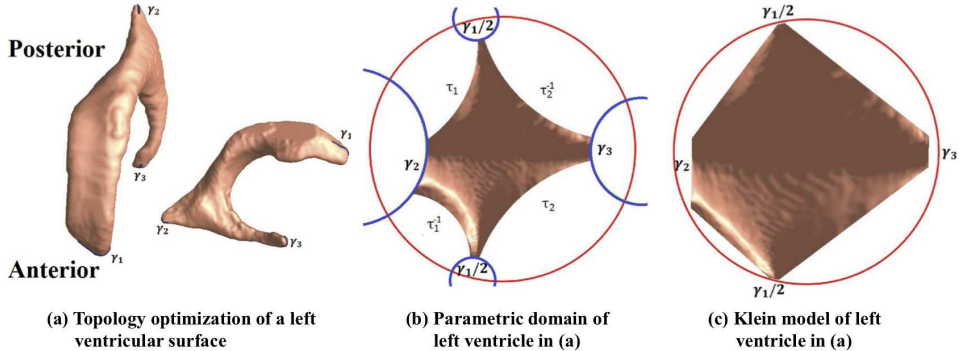}
\caption{Modeling ventricular surface with hyperbolic geometry. (a) shows three identified open boundaries, $\gamma_1$, $\gamma_2$, $\gamma_3$, on the ends of three horns. After that, ventricular surfaces can be conformally mapped to the hyperbolic space. (b) and (c) show the hyperbolic parameter space, where (b) is the Poincar\'{e} disk model and (c) is the Klein model.}
\label{fig:2}
\end{figure}

Taking a left ventricular surface $S$ as an example, the corresponding framework is summarized in Algorithm\ref{alg:HWDC}~\citep{shi2015studying} and Fig.~\ref{fig:pipeline} (c). Its critical steps are shown in Fig.\ref{fig:2}. Following our prior work~\citep{shi2015studying}, three horns of a ventricular surface are identified and three cuts $\{\gamma_1,\gamma_2,\gamma_3\}$ are made on these horns (Fig.\ref{fig:2} (a)). We term this step as \emph{topology optimization}. As a result, each ventricular surface becomes a topologically multiply connected surface and admits the hyperbolic geometry. We apply the hyperbolic Ricci flow method to compute its discrete hyperbolic uniformization metric. With the hyperbolic uniformization metric, we can embed {$S$} onto the Poincar\'e disk. In the obtained Poincar\'e disk, we apply the geodesic curve lifting algorithm~\citep{shi2015studying} to obtain a canonical parameter space (Fig.\ref{fig:2} (b)). Furthermore, we convert the Poincar\'e disk to the Klein model. It converts the canonical fundamental domains of the ventricular surfaces to a Euclidean octagon, as shown in Fig.\ref{fig:2} (c). Then we compute surface harmonic map with the Klein disk as the canonical parameter space for the following surface morphometry analysis~\citep{shi2015studying}.

\begin{algorithm}[t]
\caption{Brain surface registration with hyperbolic Ricci flow and harmonic map}
\label{alg:HWDC}
\KwIn{Brain surface $S$ with more than 2 open boundaries.}
\KwOut{Klein model of $S$}

Compute the hyperbolic uniformization metric of $S$ with hyperbolic Ricci Flow.

Compute the fundamental group of paths on $S$ and, together with original boundaries, obtain the simply connected domain $\bar{S}$.

Embed $S$ onto the Poincar\'e disk with its hyperbolic metric and its simply connected domain $\bar{S}$, we obtain the fundamental domain of $S$.

Tile the fundamental domain of $S$ with its Fuchsian group of transformations to get a finite portion of the universal covering space of $S$.

Compute the positions of the paths in the fundamental group as geodesics in the universal covering space. By slicing the universal covering space along the geodesics, we obtain the canonical fundamental domain of $S$.

Convert the canonical Poincar\'e disk to the Klein model and construct the harmonic map between $S$ and a selected template surface.


\end{algorithm}

\subsection{Surface Tensor-Based Morphometry}
\label{sec:sTBM}
Suppose $\phi = S_1 \rightarrow S_2$ is a map from surface $S_1$ to surface $S_2$. The derivative map of $\phi$ is the linear map between the tangent spaces $d\phi:TM(p)\rightarrow TM(\phi(p))$, induced by the map $\phi$, which also defines the Jacobian matrix of $\phi$. The derivative map $d\phi$ is approximated by the linear map from one face $[v_1, v_2, v_3]$ to another one $[w_1, w_2, w_3]$. First, we isometrically embed the triangles $[v_1, v_2, v_3]$ and $[w_1, w_2, w_3]$ onto the Klein disk, the planar coordinates of the vertices are denoted by $v_i, w_i, i=1, 2, 3$, which represent the 3D position of points 
Then, the Jacobian matrix for the derivative map $d\phi$ can be computed as $J=d\phi = [w_3 - w_1, w_2 - w_1][v_3 - v_1, v_2 - v_1]^{-1}.$

Based on the derivative map $J$, the surface TBM is defined as $\sqrt{det(J)}$, which measures the amount of local area changes in a surface with the map $\phi$ (Fig.~\ref{fig:pipeline} (d)). As pointed out in \citep{chung2005cortical}, each step in the processing pipeline including MRI acquisition, surface deformation, etc., are expected to introduce noise in the deformation measurement. The deformation is applied to map each subject's surface to a template surface. The Jacobian matrices of the transformation were used per subject. To account for the noise effects, we apply surface heat kernel smoothing algorithm proposed in~\citep{chung2005cortical} to improve SNR in the TBM features and boost the sensitivity of statistical analysis. The vertical-wise surface TBM features are used as the inputs for dictionary learning. We use the coordinates of these vertices to localize the ring-shaped patches and we use 3-dimentional TBM features as the feature map of ring-shape patches.

\subsection{Ring-Shaped Patch Selection}
\begin{figure}[t]
\centering
\includegraphics[height=4cm]{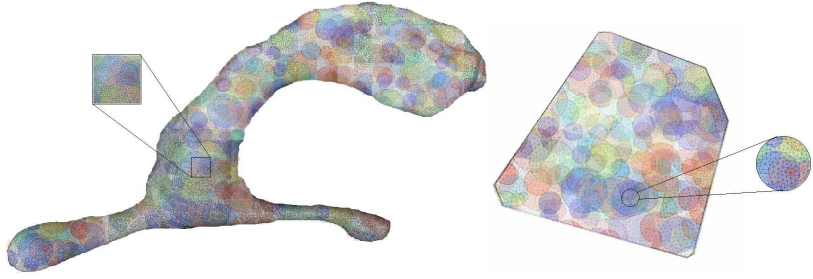}
\caption{Visualization of computed image patches on the ventricle surface (left) and hyperbolic space (right). The zoom-in pictures show some overlapping areas between image patches.}
\label{fig:3}
\end{figure}

The hyperbolic space is different from the original Euclidean space. The  common rectangle patch construction developed in Euclidean space~\citep{zhang2017multi} cannot be directly applied to the hyperbolic space. Therefore, we propose FPSBS on hyperbolic space to initialize dictionaries for sparse coding (Fig.~\ref{fig:pipeline} (e)). The intuition of the algorithm is that we want to select patches without losing the geometry information and all vertices on the hyperbolic space selected at least once. This will guarantee we learn complete information from the hyperbolic space. Fig.~\ref{fig:3} (right) is the visualization of patch selection on the hyperbolic parameter domain. And Fig.~\ref{fig:3} (left) projects the selected patches on the hyperbolic parameter domain back to the original ventricular surface, which still maintains the same topological structure as the parameter domain. In Fig.~\ref{fig:3}, each patch has a unique color and patches may overlap with each other. Together all patches cover the entire surfaces. In the following paragraph, we explain how these patches are selected.

We first randomly select a patch center point $c_1$ on the hyperbolic space $V$, where $c_1 \in V$ and $V$ is the set of all discrete vertices on the hyperbolic space. We then find all $u$ vertices connected with the center point $c_{1, i} (i = 1, 2, ..., u)$ and $c_{1, i}$ is the $i$-th vertex connected with $c_1$. The procedure is called breadth-first search (BFS)~\citep{patelcomparison}, which is an algorithm for searching graph data structures. It starts at the tree root and explores the neighbor nodes first, before moving to the next level neighbors. We use the same BFS procedure to find all connected vertices with $c_{1, i}$, which are $ c_{1,i_j} (j = 1, 2, \cdots, w_i)$. $w_i$ is the number of connected vertices with center point $c_{1, i}$. Finally, we get a vertex set (no duplicate vertices) $\mathbf{x}_1$ as follows, we call it a selected ring-shape patch on hyperbolic space and the patch center is $c_{1}$.
\begin{equation}
 \mathbf{x_1} =\{c_{1}, c_{1,1}, \cdots, c_{1,1_{w_1}}, \cdots, c_{1,u}, \cdots, c_{1,u_{w_u}} \}
\label{eqn:onepatchset}
\end{equation}

The dimension of $\mathbf{x}_1$ is $u+w_1+\cdots +w_u = m$ and $\mathbf{x}_1 \in \mathbb{R}^m$. We construct the topological patches based on hyperbolic geometry and the edge connections among different points from  $\mathbf{x}_1$. $\mathbf{x}_1$ is the first selected patch. To select the second patch center, we sample the farthest point with $c_1$, s.t. radius $r = \max_{c_{v} \in V} d_V(c_{v}, c_1)$. We now find the second patch center $c_2 \in V$ with the farthest distance $r$ of $c_1$. We follow the farthest point sampling scheme~\citep{moenning2003fast}, the sampling principle is repeatedly placing the next sample point in the middle of the least known area of the sampling domain, which can guarantee the randomness of the patch selection.

Here, $d$ is the hyperbolic distance in the Klein model. Given two points $v'$ and $v''$, draw a straight line between them; the straight line intersects the unit circle at points $a$ and $b$, so $d$ is defined as follows: 
\begin{equation}
d(v', v'')=\frac{1}{2}(\log{\frac{|av''||bv'|}{|av'||bv''|}})
\end{equation}
where $|av''|>|av'|$ and $|bv'|>|bv''|$. 

$V_r$ denotes the set of selected patch centers ($V_r=\{c_1\}$ when we compute $c_2$). After selecting the second patch $x_2$, we add $c_{2}$ into $V_r$ ($V_r=\{c_1, c_2\}$). We iterate the patch selection procedure $p$ times to get $p/2$ patches which cover every surface vertex at least once ($p/2$ patches on each side result in $p$ patches per subject). The details of FPSBS are summarized in Algorithm~\ref{alg:FBS}.

\begin{algorithm}[t]
\caption{Farthest Point Sampling with Breadth-first Search (FPSBS)}
\label{alg:FBS}
\KwIn{Hyperbolic parameter space.}
\KwOut{A collection of different amount overlapped patches on topological structure.}
Start with $V_r = \{c_{1}\}$, $V$ denotes all discrete vertices on the hyperbolic space and $V_r$ denotes the set of selected patch centers.\\
\For{T$=$1 to $n$}
{\For {$r$ determine sampling radius}
{Find set $\mathbf{x}_{T}$ by following Eq.~\ref{eqn:onepatchset} and two times BFS.\\
$r = \max_{c_{v} \in V} d_V(c_{v}, c_{T})$\\
\If{ $r \leq 10e^{-2}$ } {STOP}
\textbf{Find the farthest point $c_{T+1}$}\\
 Add $c_{T+1} = \arg\max_{c_{v} \in V}dr(c_{v}, V_r)$  to $V_r$
}
}
\end{algorithm}
\subsection{Sparse Coding and Dictionary Learning}\label{sec:sparsecoding}
We model surface TBM features as a sparse linear combination of atoms selected from a dictionary which is initialized by FPSBS on the hyperbolic parameter space. This modeling procedure is known as sparse coding~\citep{mairal2009online}. Our aim is to reduce the original surfaces dimension with the over-complete dictionary and find a linear combination of the dictionary bases to reconstruct the original surface statistics. The problem statement of sparse coding is described as below.

Given a finite training set of ring-shaped patches (as the description in Sec II. C) $\mathbf{X}=(\mathbf{x}_1, \mathbf{x}_2, \cdots, \mathbf{x}_n) \in \mathbb{R}^{m\times{n}}$, and $\mathbf{x}_{i}\in\mathbb{R}^m$, $i = 1, 2,\cdots, n$, where $m$ is the dimension of each ring-shaped patch and $n$ is the total number of patches. In this paper, we use superscript to represent $k$-th epoch and use subscript to represent $i$-th coordinate. We use boldface lower case letters $\mathbf{x}$ to denote vectors and use boldface upper case letters $\mathbf{X}$ to denote matrices. We then learn dictionary and sparse codes for these input patch features $\mathbf{x}_i$ using sparse coding.

We use $f_i(\cdot)$ to represent the optimization problem of sparse coding for each patch $\mathbf{x}_i$:
\small
\begin{equation}~\label{eqn:optimizationequation}
\min_{\mathbf{D}\in \mathbb{R}^{m \times t}, \mathbf{z}_i\in \mathbb{R}^t}{f_i}(\mathbf{D}, \mathbf{z}_i) = \frac{1}{2}||\mathbf{Dz}_i - \mathbf{x}_i||^2_2 + \lambda||\mathbf{z}_i||_1
\end{equation}
\normalsize
where $\lambda$ is the regularization parameter, $||\cdot||_2^2$ is the standard Euclidean norm and $||\mathbf{z}_i||_1 = \sum_{j=1}^{t}|z_{i, j}|$. In Eq.~\ref{eqn:optimizationequation}, each input vector will be represented by a linear combination of a few basis vectors of a dictionary. The first term of Eq.~\ref{eqn:optimizationequation} is the reconstruction error, which measures how well the new feature represents the input vector. The second term of Eq.~\ref{eqn:optimizationequation} ensures the sparsity of the learned feature $\mathbf{z}_i$. Each $\mathbf{z}_i$ is often called the \textit{sparse code}. Since $\mathbf{z}_i$ is sparse, there are only a few entries in $\mathbf{z}_i$ which are non-zero. We call its non-zero entries as its \textit{support}, i.e., supp($\mathbf{z}_i$) $= z_{i, j}: z_{i, j}\neq 0, j=1, \cdots, t$. $\mathbf{D} = (\mathbf{d}_1, \mathbf{d}_2, \cdots, \mathbf{d}_t) \in \mathbb{R}^{m\times{t}}$ is so called \textit{dictionary}, each column represents a basis vector.

\begin{figure}[t]
\centering
\includegraphics[width=4in]{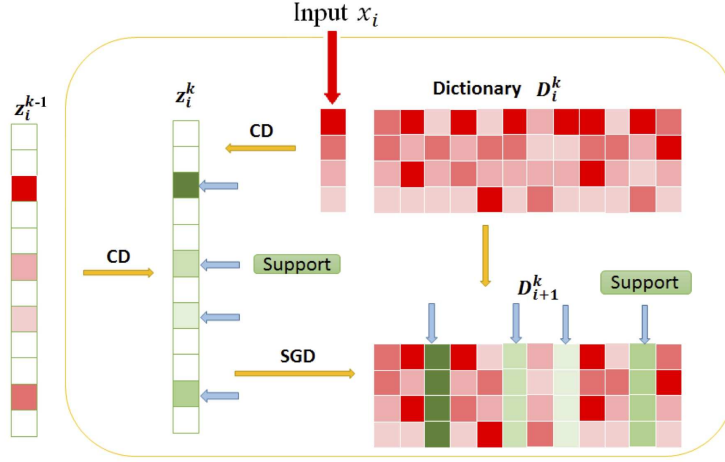}
\caption{Illustration of hyperbolic stochastic coding (HSC)  framework.}
\label{fig:alg-frame}
\vspace{-4mm}
\end{figure}

Specifically, suppose there are $t$ atoms $\mathbf{d}_j \in \mathbb{R}^m, j = 1, 2, \cdots, t$, where the number of atoms is much smaller than $n$ (the total number of training image patches) but larger than $m$ (the dimension of the image patches). $\mathbf{x}_i$ can be represented by $\mathbf{x}_i = \sum_{j=1}^{t}z_{i, j}\mathbf{d}_j$. In this way, the $m$-dimensional vector $\mathbf{x}_i$ is represented by a $t$-dimensional vector $\mathbf{z}_i = (z_{i,1}, \cdots, z_{i, t})^T$ ($\mathbf{Z} = (\mathbf{z}_1, \cdots, \mathbf{z}_n) \in \mathbb{R}^{t\times n}$). To prevent an arbitrary scaling of the sparse codes, the columns $\mathbf{d}_i$ are constrained by $\mathbb{C}\overset{\triangle}{=}\{\mathbf{D}\in \mathbb{R}^{m\times{t}}$, s.t. $\forall j = 1, \cdots, t, \mathbf{d}_j^T\mathbf{d}_j \leq 1\}$.
Thus, we use ${\cal F}(\cdot)$ to represent the sparse coding problem for $\mathbf{X}$, we then rewrite ${\cal F}(\cdot)$ as a matrix factorization problem:
\small
\begin{equation}
\min_{\mathbf{D}\in \mathbb{C}, \mathbf{Z}} {\cal F}(\mathbf{D}, \mathbf{Z})\equiv\frac{1}{n}\sum_{i=1}^{n} f_i(\mathbf{D}, \mathbf{z}_i) =\frac{1}{2}||\mathbf{X} - \mathbf{D}\mathbf{Z}||^2_F+\lambda||\mathbf{Z}||_1
\label{eq:opt}
\end{equation}
\normalsize
where $||\cdot||_F$ is the Frobenius norm. Eq. \ref{eq:opt} is a non-convex problem. However, it is a convex problem when either $\mathbf{D}$ or $\mathbf{Z}$ is fixed. When the dictionary $\mathbf{D}$ is fixed, solving each sparse code $\mathbf{z}_i$ is a Lasso problem~\citep{Tibshirani94regressionshrinkage}. Otherwise, when the $\mathbf{Z}$ are fixed, it becomes a simple quadratic problem. Here we adopt the SCC method~\citep{lin2014stochastic} to optimize Eq.~\ref{eq:opt}, which has been studied in a number of prior work~\citep[e.g.,][]{lin2014stochastic,LvSparse,lv2015holistic,Lvtask,Zhang:ISBI17,zhang2016hyperbolic,zhang2017multi,zhang2018multi}.  Following~\cite{lin2014stochastic}, we update $z_i^k$ via one or a few steps of coordinate descent (CD)~\citep{TTW08a}:

\begin{equation}\label{eq:cd-obj-a}
\mathbf{z}_i^k = \mathrm{CD}( \mathbf{D}^k_i, \mathbf{z}_i^{k-1},\mathbf{x}_i )
\end{equation}
The updated sparse code is then denoted by $\mathbf{z}^{k}_i$. A detailed derivation of CD utilizing software thresholding shrinkage function~\citep{Combettes2005-MMS} can be found in \cite{lin2014stochastic}. We then update the dictionary $\mathbf{D}$ by using stochastic gradient descent (SGD)~\citep{bottou1998online}:
\begin{equation}\label{eq:sgd}
\mathbf{D}^k_{i+1} = P_{\mathbb{C}}(\mathbf{D}_{i}^k - \eta_{i}^k \nabla_{\mathbf{D}_{i}^k} f_i(\mathbf{D}_{i}^k, \mathbf{z}_i^k))
\end{equation}
where $P$ is the shrinkage function, $\mathbb{C}$ is the feasible set of $\mathbf{D}$ and $\eta_i^k$ is the learning rate of $i$-th step in $k$-th epoch. We set the learning rate as an approximation of the inverse of the Hessian matrix $\mathbf{H}$. We illustrate the algorithmic framework in Fig.~\ref{fig:alg-frame}. At each iteration, with a ring-shaped patch $\mathbf{x}_i$, we perform one step of CD to find the supports of the sparse code $\mathbf{z}_i^{k-1}$. Next, we perform a few steps of CD on the supports to obtain a new sparse code $\mathbf{z}_i^{k}$. Then we update the supports of the dictionary by the second order SGD to obtain a new dictionary $\mathbf{D}^k_{i+1}$.

\subsection{Max-Pooling and Adaboost Classifier}\label{sec:maxpooling}
With a trained dictionary $\mathbf{D}$, for a set of ring-shaped patches from a new subject, $\mathbf{x}_i$, $i=1, 2, ..., p$, $p$ is the patch number of an individual subject, we can learn its sparse features $\mathbf{z}_i$, $\mathbf{z_i}\in \mathbb{R}^t, i=1, 2, ..., p$. In theory, one could use all learned features as input data of a classifier but it poses intractable computational challenges. Thus, to describe our surface features efficiently, one natural approach is to aggregate statistics of these features at various locations. A key component of deep learning models, max-pooling~\citep{boureau2010theoretical} takes the most responsive node of a given region of interest. In our system, we borrow the idea of max-pooling and apply it to the extracted sparse coding surface features (sparse codes) from HSC (Fig.~\ref{fig:pipeline} (g)). Specifically, one could compute the max value for each feature ($t$ features obtained from HSC) over all patches ($p$ patches per subject), which is equivalent to applying a high-pass filter to the learned sparse codes. These summary statistics are much lower in dimension $t$ compared to using all of the learned surface patch features and reduce over fitting. Finally, Adaboost~\citep{rojas2009adaboost} classifier is used for binary classification, as shown in Fig.~\ref{fig:pipeline} (h).

\section{Results}

\subsection{Dataset I: MCI Converter vs. MCI stable Subjects}

In Dataset I, we try to use ventricular morphometry features to discriminate between MCIc and MCIs subjects. To extract hyperbolic surface features, we automatically segment lateral ventricular volumes with the multi-atlas fluid image alignment (MAFIA) method~\citep{chou2010ventricular} from each MRI scan. We then use a topology-preserving level set method~\citep{han2009a} to build surface models and the marching cube algorithm~\citep{lorensen1987marching} is applied to construct triangular surface meshes (Fig.~\ref{fig:pipeline} (b)). After the topology optimization, we apply hyperbolic Ricci flow method and conformally map the ventricular surface to the Poincar\'{e} disk and register them via harmonic map~\citep{shi2015studying}. We finally compute the surface TBM features~\citep{shi2015studying} and smooth them with surface heat kernel method~\citep{chung2005cortical}.

Here we select $2,000$ ring-shaped patches (300-vertex, Fig.~\ref{fig:3}) by the FPSBS method (Alg.~\ref{alg:FBS}) on each side of ventricle for each subject and finally have $n = 532, 000$ ring-shaped patches. The generated patches are consistent across all subjects since  surfaces are registered already~\citep{shi2015studying}. Surface TBM is a scalar feature defined on each vertex on the hyperbolic domain so the feature number of each subject is $1,200,000$ ($m=300$, $p=4,000$ in notations of Sec.~\ref{sec:sparsecoding},~\ref{sec:maxpooling}). We initialize the dictionary via selecting random patches~\citep{coates2011importance}, which has been shown to be a very efficient method in practice. Then we learn the dictionary and sparse codes by HSC using the initial dictionary. All our experiments involve training for ten epochs with a batch size of one. After the optimization, all subjects use the same dictionary ($m=300$, $t=2,000$ in notation of Sec.~\ref{sec:sparsecoding}). With the sparse coding, we obtain $4,000$ samples each of which has $2,000$ sparse codes per subject. After that, max-pooling is adopted to choose the maximum value for each sparse code as a feature on these $4,000$ samples. Our final feature dimension for classification is $2,000$ per subject.

We take a nested cross-validation approach by pre-separating training, validation and testing sets. Specifically, we use the ratio of 7:1:2 for training, validation and testing.  We select the hyper parameters based on the validation set and test all methods on the same testing set.
Besides, we also compare our work with some other measures and methods. We compute bilateral ventricular volumes and surface areas, which are used as MRI biomarkers in AD research. We also compare HSC with a ventricular surface shape method in~\citep{pmid18228600} (\emph{Shape}), which automatically generates comparable meshes of all ventricles. The deformations based on the morphometry model are employed with repeated permutation tests and then used as geometry features. With our ventricle surface registration results, we follow the \emph{Shape} work~\citep{pmid18228600} for selecting biomarkers and use support vector machine (SVM) for classification on the same dataset. We implement the low-rank shared dictionary learning (LRSDL) method, based on the paper~\citep{vu2017fast} and the github source code~\footnote{ https://github.com/tiepvupsu/DICTOL\_python}. We select the hyper-parameter for LRSDL by using the same strategy as HSC on the training set. We run LRSDL 50 iterations with $\lambda_1=0.13$, $\lambda_2=0.1$, $\eta=0.05$, $k=20$, $k_0=10$. Same as HSC, we apply the LRSDL on ring-shape patches and apply max-pooling as post processing on the learnt sparse codes. The same classifier is applied on the learnt features on the same test set as HSC. We test \emph{HSC, Shape, volume, area and LRSDL} measures on the left, right and whole ventricle, respectively. Accuracy (ACC), Sensitivity (SEN), Specificity (SPE) and compute Area Under The Curve (AUC) are computed to evaluate classification results. Table~\ref{tab:4} shows classification performances of four methods.
\begin{table}[t]
\centering

\begin{tabular}{ccccccc}
\hline
Dataset I              &     & HSC     & Shape    & Volume  & Area    & LRSDL   \\ \hline
\multirow{4}{*}{whole} & ACC & 85.19\% & 70.37\%  & 66.67\% & 59.26\% & 77.78\% \\
                       & SEN & 76.92\% & 100.00\% & 85.71\% & 78.57\% & 75.00\% \\
                       & SPE & 81.25\% & 57.89\%  & 63.16\% & 57.89\% & 80.00\% \\
                       & AUC & 0.8516 & 0.7857  & 0.6731 & 0.5934 & 0.7775 \\
                       \hline
\multirow{4}{*}{left}  & ACC & 74.07\% & 81.48\%  & 62.96\% & 59.26\% & 70.37\% \\
                       & SEN & 76.92\% & 76.92\%  & 61.54\% & 83.33\% & 76.92\% \\
                       & SPE & 71.43\% & 75.00\%  & 56.52\% & 52.63\% & 64.29\% \\
                      & AUC & 0.7418 & 0.8159  & 0.6264 & 0.5907 & 0.706 \\
                      \hline
\multirow{4}{*}{right} & ACC & 70.37\% & 66.67\%  & 62.96\% & 59.26\% & 70.37\% \\
                       & SEN & 53.85\% & 84.62\%  & 84.62\% & 78.57\% & 69.23\% \\
                       & SPE & 78.57\% & 57.14\%  & 57.89\% & 57.89\% & 71.43\% \\
                       & AUC & 0.7088 & 0.6703  & 0.6319 & 0.5879 & 0.7033 \\
                      \hline
\end{tabular}
\caption{The comparison results by nested cross-validation on Dataset I.}\label{tab:4}
\end{table}


From the experimental results, we can find that the best accuracy (85.19\%) and the best specificity (81.25\%) are achieved when we use TBM features on ventricle hyperbolic space of both sides (whole) for training and testing. The comparison shows that our new framework selects better features, and achieves better and more meaningful classification results. 
The HSC algorithm with whole ventricle TBM features achieves the best AUC (0.8516). The comparison demonstrates that our proposed algorithm may be useful for AD diagnosis and prognosis research.

\subsection{Dataset II: ADNI Baseline Cortical Surfaces}
\label{data1}
\begin{figure}[t]
\centering
\includegraphics[width=5in]{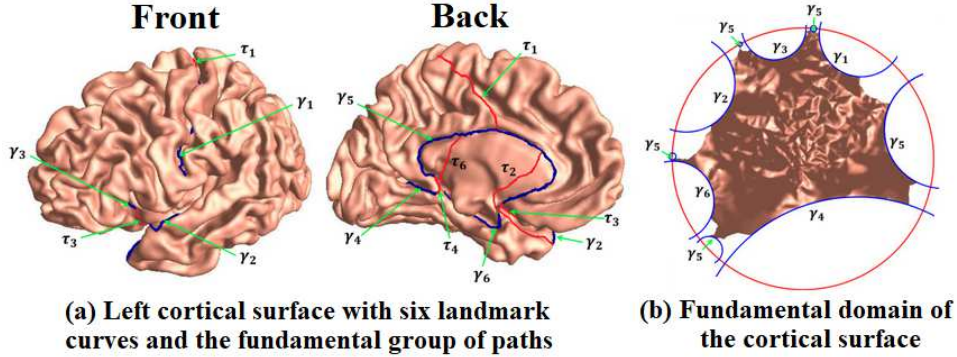}
\caption{Modeling cortical surface with hyperbolic geometry. (a) shows six identified open boundaries, $\gamma_1, \cdots, \gamma_6$. (b) shows the hyperbolic parameter space, which is the Poincar\'{e} disk model}
\label{fig:7}
\end{figure}
Many researches have analyzed that the cortial surface morphometry is a valid imaging biomarker for  AD~\citep{shi2019hyperbolic,thompson2004mapping,chung2008tensor}. In Dataset II, we apply HSC to analyze cortical morphometry for AD related clinical group classification. We use the left hemispheric cerebral cortices and follow~\citep{shi2019hyperbolic} to preprocess cortical surface data. We first use FreeSurfer software~\citep{fischl2012freesurfer} to preprocess the MRIs of 115 subjects and reconstruct their left cortical surfaces. The Caret software~\citep{van2012cortical} is then used to automatically label six major brain landmarks, which include the Central Sulcus, Anterior Half of the Superior Temporal Gyrus, Sylvian Fissure, Calcarine Sulcus, Medial Wall Ventral Segment and Medial Wall Dorsal Segment. Fig.~\ref{fig:7} (a) shows an example of the landmark curves on the left cortical surface, where the six landmark curves are modeled as open boundaries and denoted as $\gamma_1, \cdots, \gamma_6$. The fundamental group of paths are computed by connecting boundary $\gamma_5$ to every other boundary and the path is denoted as $\tau_1, \tau_2, \tau_3, \tau_4, \tau_6$. Fig.~\ref{fig:7} (b) shows that they are embedded into the Poincar\'{e} disk. After we cut the cortical surfaces along the delineated landmark curves, the cortical surfaces become genus-0 surfaces with six open boundaries. We finally randomly select the left cortical surface of a CU subject, who is not in the studied subject dataset, as the template surface, and perform the processing steps described in Sec.~\ref{sec:HRF} and Sec.~\ref{sec:sTBM} to get the hyperbolic surface TBM features.

\begin{figure}[t]
\centering
\includegraphics[height=4cm]{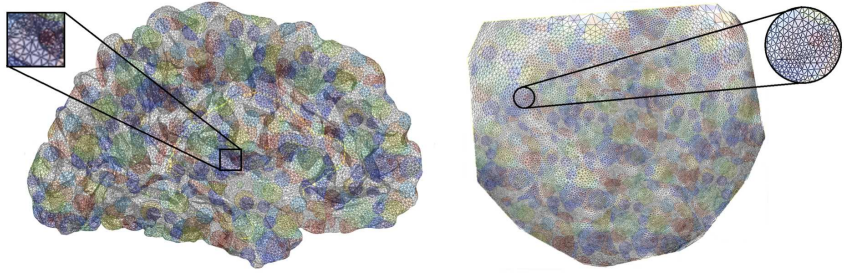}
\caption{Visualization of computed image patches on the cortical surface (left) and hyperbolic space (right). Each patch has a unique color. The zoom-in pictures show some overlapping areas between image patches.}
\label{fig:8}
\end{figure}

For Dataset II, since there are only 115 subject for three classes, we use the same hyperparameter as what we used in Dataset I for training. We apply five-fold cross-validation to evaluate our algorithm, which guarantees the model is tested on all subjects.
All experiments are trained for $k= 10$ epochs with a batch size of 1. The regularization parameter $\lambda$ is set to $0.10 \approx 1.2/\sqrt{m}$, $1/\sqrt{m}$ is a classical normalization factor~\citep{bickel2009simultaneous} and the constant 1.2 has been shown to produce about 10 non-zero coefficients. We select $p = 2,000$ ring-shaped patches as shown in Fig.~\ref{fig:8} by FPSBS on the cortical surface and we have $n = 230,000$ ring-shaped patches for Dataset II. Fig.~\ref{fig:8} (right) is the visualization of cortical morphometry on the hyperbolic parameter domain and Fig.~\ref{fig:8} (left) projects the selected patches on the hyperbolic parameter domain back to the original cortical surface. Our FPSBS patch selection algorithm can maintain the same topological structure as the parameter domain.
After learning the sparse codes via HSC, we apply max-pooling~\citep{boureau2010theoretical} for further dimension reduction. Finally, we employ the Adaboost~\citep{rojas2009adaboost} to do the binary classification and distinguish individuals from different groups. We report the classification results of (1) AD vs. CU, (2) AD vs. MCI and (3) MCI vs. CU in Table~\ref{tab:3}. 

\begin{table}[t]
\centering
\begin{tabular}{cccc}
\hline
Dataset II & AD vs. CU & AD vs. MCI & MCI vs. CU \\ \hline
ACC       & 88.57\%   & 82.67\%    & 80\%       \\
SEN       & 89.29\%   & 84.00\%    & 79.17\%    \\
SPE       & 83.33\%   & 70.00\%    & 84.44\%    \\ 
AUC       & 0.879   & 0.8111    & 0.7944    \\ \hline
\end{tabular}
\caption{The classification results by five-fold cross-validation on Dataset II.}\label{tab:3}
\end{table}
In our prior work~\citep{shi2019hyperbolic}, we have shown that the hyperbolic surface features are significantly associated with the diagnostic disease severity.  However, it is difficult to directly use hyperbolic surface features for different stages of disease diagnosis classification due to the large amount of features and limited subject numbers. Table~\ref{tab:3} shows that HSC overcomes the above issue and FPSBS has a good generalization capability to capture the meaningful features from ring-shaped patches. HSC works well on even more subtle difference classification problem (CU vs. MCI) compared with AD vs. CU. Our new framework makes meaningful and high performances on different groups and may be useful for AD diagnosis and prognosis researches.





\section{Discussion}
The current work presents our initial efforts to develop efficient machine learning methods to work with brain sMRI features computed from general topological surfaces. We validate our proposed FPSBS and HSC methods on two datasets and the preliminary experimental results demonstrate that the proposed algorithms outperform some other works on 
classification accuracy. By reducing the dimension of hyperbolic TBM features with the novel HSC algorithm, the present study is capable of applying the low-dimensional HSC measures to diagnose AD and its prodromal stages. In Dataset I, the proposed system successfully distinguishes the ventricular HSC measures of MCIc subjects from MCIs subjects with a higher accuracy ($>85\%$) than the classification systems using whole ventricular volume, area and surface-based biomarkers. 
In Dataset II, the proposed system has an outstanding performance to discriminate the cortical HSC measures of AD, MCI and CU groups (accuracy$>80\%$).  
These experimental results are consistent with our hypothesis that the lower-dimensional TBM statistics (or HSC measures) may outperform volume, area and shape-based structural measures on discriminating kinds of symptomatic groups related with AD. 

Our study has two main findings. First, in order to better initialize the dictionary for sparse coding on the hyperbolic parameter domain, we propose a ring-shaped patch selection algorithm -- FPSBS -- to capture the surface features. The extracted patch structures help reduce feature noises and enhance statistical power of the computed surface TBM features. Second, we propose an efficient sparse coding and dictionary framework defined in the hyperbolic parameter space -- HSC. To the best of our knowledge, HSC is the first sparse coding framework which is designed for general topological surfaces admitting the hyperbolic geometry. Our work may shed new lights on how to extract important surface features from hyperbolic geometry analyses.

In this paper, similar to our prior work on Euclidean parameter domain~\citep{Zhang:ISBI2016}  and spherical parameter domain~\citep{zhang2017empowering} , we propose patch-based sparse coding feature selection method to leverage the surface analysis for early disease diagnosis.  Differently, our current ring-shaped patch selection algorithm, farthest point sampling with breadth-first search (FPSBS) method, captures the surface features within the hyperbolic space. The proposed FPSBS method can better capture surface features to initialize the dictionary for sparse coding. It is worth noting that the FPSBS is a general algorithm and can also be applied to any Riemann surfaces. The current application to surfaces with general topology is remarkable because, different from Euclidean and spherical parameter domains, it is difficult to generate regular grids on the hyperbolic parameter domain. Our method provides a convenient way to take advantage of the intrinsic geometry structure of hyperbolic parameter domain and generate efficient patch coverage for surfaces with general topology. Therefore, the ring shaped patch selection is an important component of the current brain image analysis system.

We present an efficient sparse coding and dictionary learning method which works on features extracted from surfaces with general topology admitting the hyperbolic geometry. The work is a generalization and enrichment of existing SCC research~\citep{lin2014stochastic,LvSparse,lv2015holistic,Lvtask,zhang2016hyperbolic,zhang2017multi,zhang2018multi,Zhang:ISBI17}. Overall,  our work show that the proposed HSC 
may extract critical shape features for classifying different stages of AD. Our work should be very useful and meaningful in brain image analysis, shape analysis and machine learning research. HSC has the potential to be a useful tool to study 3D surfaces with general topology structures.

To identify patients as early as possible in the course of MCI, ADNI subdivided MCI into ``early'' and ``late'' stages based on the Wechsler Memory Scale-Revised (WMS-R) Logical Memory II Story A score~\citep{petersen2010alzheimer}. Studies under this MCI staging framework reveal distinguished biomarkers of “early” MCIs (EMCL) as compared to other clinical groups~\citep{wee2019cortical,tosun2014multimodal}. We also applied our proposed method on EMCI and “late” MCI (LMCI) in our recent work, and achieved 84\% accuracy, on ADNI2-dataset with 37 LMCI and 73 EMCI~\citep{zhang2017empowering}. However, studies of~\citep{brooks2007substantial,brooks2008potential,de2005accidental} show that it is overinclusive to use this ADNI staging method based on a single test score, the study of~\citep{edmonds2019early} revealed that ADNI’s early MCI group included a large proportion (56\%) of false-positive diagnostic errors. 

Another important problem in AD diagnosis research is to identify the probability of MCI patients converting to dementia. Identifying MCI converters from the MCI stable is critical for clinical management. Based on MRI data,~\cite{shi2015studying} successfully revealed significant ventricular morphometry differences between MCI converter group and MCI stable group using a novel ventricular morphometry analysis system. Using the same MRI cohorts, this work further proposes HSC to extract effective structural features and classify 71 MCI converters versus 62 nonconverters. The AUC is 0.85.~\cite{moradi2015machine} proposed a low density separation scheme to learn aggregate biomarkers to discriminate 164 MCI converters from 100 stable MCI patients. They achieved a 0.7661 AUC. The study of~\citep{sorensen2016early} applied hippocampal texture features and support vector machine (SVM) classifier to predict 8 MCI converters versus 17 nonconverters. The best prognostic AUC was 0.83.~\cite{chincarini2011local} applied a medial temporal lobe intensity and textural features and SVM classifier to separate 136 MCI converters from 166 MCI non converters with AUC=0.74.~\cite{collij2016application} applied SVM classifier to predict arterial spin labeling perfusion maps of 12 patients with MCI diagnosis converted to AD versus 12 subjects with stable MCI. The AUC was 0.71. Table~\ref{tab:5} presents the AUC values of this work and the above studies. Compared to other single modality neuroimaging-based classifiers, our proposed system has a larger or comparable AUC on predicting MCI converters versus nonconverters. There are also studies demonstrating that multimodality machine learning models have superior performances than single modality classifiers~\citep{varatharajah2019predicting,rathore2017review,moradi2015machine}. We have developed a series of surface-based biomarkers of various brain structures for AD research~\citep{dong2020applying,Wang:NIMG10,wang2011surface,dong2019applying,fan2018tetrahedron}. Our latest work~\citep{dongintegrating} has indicated that combining these biomarkers could empower the prediction of AD progression. In future work, we expect to improve the MCI conversion prediction performance by introducing these effective multiple statistics. We also note that the study~\citep{edmonds2019early} proposed a neuropsychological approach to improve the reliability of staging early and late MCI. We expect that our biomarkers for predicting MCI converters versus nonconverters also work well with reliable neuropsychological tests for staging early and late MCI. We will study it in our future work.
\begin{table}[t]
\centering
\begin{tabular}{cccc}
\hline
Method & Subjects (MCIc/MCInc) & Feature & AUC \\ \hline
HSC (this paper)     & 71/62  & \begin{tabular}{@{}c@{}}Ventricular surface\\ TBM features\end{tabular}   & 0.85       \\
\begin{tabular}{@{}c@{}}Low density separation\\  ~\citep{moradi2015machine} \end{tabular} & 164/100 & Gray matter density  & 0.77    \\
\begin{tabular}{@{}c@{}}Support vector machine (SVM)\\~\citep{sorensen2016early}  \end{tabular}& 8/17   & Hippocampal texture    & 0.83    \\
SVM~\citep{chincarini2011local} & 136/166   &\begin{tabular}{@{}c@{}} Medial temporal lobe\\ intensity and textural\\ MRI-based features\end{tabular}    & 0.74   \\ 
SVM~\citep{collij2016application} & 12/12   &\begin{tabular}{@{}c@{}} Arterial spin labeling\\ perfusion maps\end{tabular}    & 0.71   \\ \hline
\end{tabular}
\caption{Studies to distinguish MCI converters (MCIc) from nonconverters (MCInc).}\label{tab:5}
\end{table}

There are three important caveats when applying the proposed framework to AD diagnosis and prognosis. First, because of the overlapping patch selection and max-pooling scheme, we generally cannot visualize the selected features and it decreases the comprehensibility although we may always visualize statistically significant regions in our prior group difference studies~\citep{shi2013surface,wang2013applying}. However, our recent work~\citep{zhang2018multi} made some progress which may potentially better address this problem. In our recent work~\citep{zhang2018multi}, instead of randomly selecting patches to build the initial dictionary, we use group lasso screening to select the most significant features. Therefore, the features used in sparse coding may be visualized on the surface map. In future, we will incorporate this idea into our current framework to improve its interpretation ability. Second, our current work, similar to several other work~\citep[e.g.][]{fan2007compare,colliot2008discrimination,kloppel2008automatic,gerardin2009multidimensional,magnin2009support,cuingnet2011automatic,liu2011combination,shen2012detecting,ahmed2015classification}, uses clinical diagnoses as the ``ground truth'' diagnoses for training and cross-validation. However, some recent work~\citep[e.g.][]{Beach:JNEN12} has reported that neuropathological diagnoses only had limited accuracy values (e.g. only 80 - 90\% of the labels are correct) when confirmed with AD histopathology. Under this limitation, we should be cautious when making inferences and conclusions on our work for the AD diagnosis since our discovered features are not necessarily real AD biomarkers. Even so, our recent work~\citep{wu2018hippocampus} has studied hippocampal morphometry on a cohort consisting of $A\beta$ positive AD ($N=151$) and matched $A\beta$ negative cognitively unimpaired subjects ($N=271$) where $A\beta$ positivity was determined using mean-cortical standard uptake value ratio (SUVR) with cerebellum as the reference region over the amyloid PET images. With our Euclidean SCC work~\citep{Zhang:ISBI2016} integrating the proposed HSC and MP methods, we achieved an accuracy rate of 90.48\% in that task~\citep{wu2018hippocampus}. The results demonstrate that our proposed framework may potentially help discover pathology-confirmed AD biomarkers. Third, as our initial attempt to integrate geometry analysis and machine learning method for AD diagnosis, the current work reports very limited  experimental results since we mainly reuse the data in our prior published work~\citep{shi2015studying,shi2019hyperbolic}. The geometry analysis part involves multiple steps, including image segmentation, surface registration, surface parameterization, etc. Our ongoing work, e.g.~\citep{Mi:AAAI20}, is developing novel approaches which will make the whole process more automatic and more accurate. Once they are available, we will apply the proposed method to analyze more longitudinal ADNI data.

\section{Conclusions and Future Works}
Here we present a hyperbolic sparse coding with ring-shaped patch selection algorithm, which may improve the 
accuracy for AD diagnosis and prognosis with sMRI biomarkers. In the future, we will explore whether the proposed framework will work with other shape statistics, such as spherical harmonics and radial distance. In our previous preclinical AD study~\citep{dong2019applying}, we found APOE-e4 does effects on hippocampal morphometry of cognitively unimpaired subjects. In our future work, we will further explore whether the proposed FPSBS and HSC methods are useful for such preclinical AD study.  

\begin{center}
{\large Acknowledgements}    
\end{center}

This research is supported in part by National Institutes of Health (RF1AG051710, R21AG065942, R01EB025032, U54EB020403, R01AG031581 and P30AG19610), National Science Foundation (IIS-1421165) and Arizona Alzheimer's Consortium.

Data collection and sharing for this project was funded by the Alzheimer’s Disease Neuroimaging Initiative (ADNI) (National Institutes of Health Grant U01 AG024904) and DOD ADNI (Department of Defense award number W81XWH-12-2-0012). ADNI is funded by the National Institute on Aging, the National Institute of Biomedical Imaging and Bioengineering, and through generous contributions from the following: AbbVie, Alzheimers Association; Alzheimers Drug Discovery Foundation; Araclon Biotech; BioClinica, Inc.; Biogen; Bristol-Myers Squibb Company; CereSpir, Inc.; Cogstate; Eisai Inc.; Elan Pharmaceuticals, Inc.; Eli Lilly and Company; EuroImmun; F. Hoffmann-La Roche Ltd and its affiliated company Genentech, Inc.; Fujirebio; GE Healthcare; IXICO Ltd.; Janssen Alzheimer Immunotherapy Research \& Development, LLC.; Johnson \& Johnson harmaceutical Research \& Development LLC.; Lumosity; Lundbeck; Merck \& Co., Inc.; Meso Scale Diagnostics, LLC.; NeuroRx Research; Neurotrack Technologies; Novartis Pharmaceuticals Corporation; Pfizer Inc.; Piramal Imaging; Servier; Takeda Pharmaceutical Company; and Transition Therapeutics. The Canadian Institutes of Health Research is providing funds to support ADNI clinical sites in Canada. Private sector contributions are facilitated by the Foundation for the National Institutes of Health (www.fnih.org). The grantee organization is the Northern California Institute for Research and Education, and the study is coordinated by the Alzheimers Therapeutic Research Institute at the University of Southern California. ADNI data are disseminated by the Laboratory for NeuroImaging at the University of Southern California.

\begin{center}
{\large References}
\end{center}

\bibliography{fbs}

\begin{thebibliography}{105}
\expandafter\ifx\csname natexlab\endcsname\relax\def\natexlab#1{#1}\fi
\providecommand{\url}[1]{\texttt{#1}}
\providecommand{\href}[2]{#2}
\providecommand{\path}[1]{#1}
\providecommand{\DOIprefix}{doi:}
\providecommand{\ArXivprefix}{arXiv:}
\providecommand{\URLprefix}{URL: }
\providecommand{\Pubmedprefix}{pmid:}
\providecommand{\doi}[1]{\href{http://dx.doi.org/#1}{\path{#1}}}
\providecommand{\Pubmed}[1]{\href{pmid:#1}{\path{#1}}}
\providecommand{\bibinfo}[2]{#2}
\ifx\xfnm\relax \def\xfnm[#1]{\unskip,\space#1}\fi
\bibitem[{Beach et~al.(2012)Beach, Monsell, Phillips and Kukull}]{Beach:JNEN12}
\bibinfo{author}{Beach, T.G.}, \bibinfo{author}{Monsell, S.E.},
  \bibinfo{author}{Phillips, L.E.}, \bibinfo{author}{Kukull, W.},
  \bibinfo{year}{2012}.
\newblock \bibinfo{title}{{{A}ccuracy of the clinical diagnosis of {A}lzheimer
  disease at {N}ational {I}nstitute on {A}ging {A}lzheimer {D}isease {C}enters,
  2005-2010}}.
\newblock \bibinfo{journal}{J. Neuropathol. Exp. Neurol.} \bibinfo{volume}{71},
  \bibinfo{pages}{266--273}.
\bibitem[{Ben~Ahmed et~al.(2015)Ben~Ahmed, Mizotin, Benois-Pineau, Allard,
  Catheline and Ben~Amar}]{ahmed2015classification}
\bibinfo{author}{Ben~Ahmed, O.}, \bibinfo{author}{Mizotin, M.},
  \bibinfo{author}{Benois-Pineau, J.}, \bibinfo{author}{Allard, M.},
  \bibinfo{author}{Catheline, G.}, \bibinfo{author}{Ben~Amar, C.},
  \bibinfo{year}{2015}.
\newblock \bibinfo{title}{{{A}lzheimer's disease diagnosis on structural {M}{R}
  images using circular harmonic functions descriptors on hippocampus and
  posterior cingulate cortex}}.
\newblock \bibinfo{journal}{Comput Med Imaging Graph} \bibinfo{volume}{44},
  \bibinfo{pages}{13--25}.
\bibitem[{Bickel et~al.(2009)Bickel, Ritov and
  Tsybakov}]{bickel2009simultaneous}
\bibinfo{author}{Bickel, P.J.}, \bibinfo{author}{Ritov, Y.},
  \bibinfo{author}{Tsybakov, A.B.}, \bibinfo{year}{2009}.
\newblock \bibinfo{title}{Simultaneous analysis of lasso and dantzig selector}.
\newblock \bibinfo{journal}{The Annals of Statistics} ,
  \bibinfo{pages}{1705--1732}.
\bibitem[{Bottou(1998)}]{bottou1998online}
\bibinfo{author}{Bottou, L.}, \bibinfo{year}{1998}.
\newblock \bibinfo{title}{Online learning and stochastic approximation}.
\newblock \bibinfo{journal}{Online Learning and Neural Networks, Cambridge
  University Press, Cambridge, UK} .
\bibitem[{Boureau et~al.(2010)Boureau, Ponce and
  LeCun}]{boureau2010theoretical}
\bibinfo{author}{Boureau, Y.L.}, \bibinfo{author}{Ponce, J.},
  \bibinfo{author}{LeCun, Y.}, \bibinfo{year}{2010}.
\newblock \bibinfo{title}{A theoretical analysis of feature pooling in visual
  recognition}, in: \bibinfo{booktitle}{Proceedings of the 27th international
  conference on machine learning (ICML-10)}, pp. \bibinfo{pages}{111--118}.
\bibitem[{Brooks et~al.(2008)Brooks, Iverson, Holdnack and
  Feldman}]{brooks2008potential}
\bibinfo{author}{Brooks, B.L.}, \bibinfo{author}{Iverson, G.L.},
  \bibinfo{author}{Holdnack, J.A.}, \bibinfo{author}{Feldman, H.H.},
  \bibinfo{year}{2008}.
\newblock \bibinfo{title}{Potential for misclassification of mild cognitive
  impairment: A study of memory scores on the {Wechsler Memory Scale-III} in
  healthy older adults}.
\newblock \bibinfo{journal}{Journal of the International Neuropsychological
  Society} \bibinfo{volume}{14}, \bibinfo{pages}{463--478}.
\bibitem[{Brooks et~al.(2007)Brooks, Iverson and White}]{brooks2007substantial}
\bibinfo{author}{Brooks, B.L.}, \bibinfo{author}{Iverson, G.L.},
  \bibinfo{author}{White, T.}, \bibinfo{year}{2007}.
\newblock \bibinfo{title}{Substantial risk of {“Accidental MCI”} in healthy
  older adults: Base rates of low memory scores in neuropsychological
  assessment}.
\newblock \bibinfo{journal}{Journal of the International Neuropsychological
  Society} \bibinfo{volume}{13}, \bibinfo{pages}{490--500}.
\bibitem[{Chincarini et~al.(2011)Chincarini, Bosco, Calvini, Gemme, Esposito,
  Olivieri, Rei, Squarcia, Rodriguez, Bellotti et~al.}]{chincarini2011local}
\bibinfo{author}{Chincarini, A.}, \bibinfo{author}{Bosco, P.},
  \bibinfo{author}{Calvini, P.}, \bibinfo{author}{Gemme, G.},
  \bibinfo{author}{Esposito, M.}, \bibinfo{author}{Olivieri, C.},
  \bibinfo{author}{Rei, L.}, \bibinfo{author}{Squarcia, S.},
  \bibinfo{author}{Rodriguez, G.}, \bibinfo{author}{Bellotti, R.}, et~al.,
  \bibinfo{year}{2011}.
\newblock \bibinfo{title}{Local {MRI} analysis approach in the diagnosis of
  early and prodromal {A}lzheimer's disease}.
\newblock \bibinfo{journal}{Neuroimage} \bibinfo{volume}{58},
  \bibinfo{pages}{469--480}.
\bibitem[{Chou et~al.(2010)Chou, Lepor{\'e}, Saharan, Madsen, Hua, Jack, Shaw,
  Trojanowski, Weiner, Toga et~al.}]{chou2010ventricular}
\bibinfo{author}{Chou, Y.Y.}, \bibinfo{author}{Lepor{\'e}, N.},
  \bibinfo{author}{Saharan, P.}, \bibinfo{author}{Madsen, S.K.},
  \bibinfo{author}{Hua, X.}, \bibinfo{author}{Jack, C.R.},
  \bibinfo{author}{Shaw, L.M.}, \bibinfo{author}{Trojanowski, J.Q.},
  \bibinfo{author}{Weiner, M.W.}, \bibinfo{author}{Toga, A.W.}, et~al.,
  \bibinfo{year}{2010}.
\newblock \bibinfo{title}{Ventricular maps in 804 {ADNI} subjects: correlations
  with {CSF} biomarkers and clinical decline}.
\newblock \bibinfo{journal}{Neurobiology of aging} \bibinfo{volume}{31},
  \bibinfo{pages}{1386--1400}.
\bibitem[{Chung et~al.(2008)Chung, Dalton and Davidson}]{chung2008tensor}
\bibinfo{author}{Chung, M.K.}, \bibinfo{author}{Dalton, K.M.},
  \bibinfo{author}{Davidson, R.J.}, \bibinfo{year}{2008}.
\newblock \bibinfo{title}{Tensor-based cortical surface morphometry via
  weighted spherical harmonic representation}.
\newblock \bibinfo{journal}{Medical Imaging, IEEE Transactions on}
  \bibinfo{volume}{27}, \bibinfo{pages}{1143--1151}.
\bibitem[{Chung et~al.(2005)Chung, Robbins, Dalton, Davidson, Alexander and
  Evans}]{chung2005cortical}
\bibinfo{author}{Chung, M.K.}, \bibinfo{author}{Robbins, S.M.},
  \bibinfo{author}{Dalton, K.M.}, \bibinfo{author}{Davidson, R.J.},
  \bibinfo{author}{Alexander, A.L.}, \bibinfo{author}{Evans, A.C.},
  \bibinfo{year}{2005}.
\newblock \bibinfo{title}{Cortical thickness analysis in autism with heat
  kernel smoothing}.
\newblock \bibinfo{journal}{NeuroImage} \bibinfo{volume}{25},
  \bibinfo{pages}{1256--1265}.
\bibitem[{Coates and Ng(2011)}]{coates2011importance}
\bibinfo{author}{Coates, A.}, \bibinfo{author}{Ng, A.Y.}, \bibinfo{year}{2011}.
\newblock \bibinfo{title}{The importance of encoding versus training with
  sparse coding and vector quantization}, in: \bibinfo{booktitle}{Proceedings
  of the 28th International Conference on Machine Learning (ICML-11)}, pp.
  \bibinfo{pages}{921--928}.
\bibitem[{Collij et~al.(2016)Collij, Heeman, Kuijer, Ossenkoppele, Benedictus,
  M{\"o}ller, Verfaillie, Sanz-Arigita, van Berckel, van~der Flier, ,
  Scheltens, Barkhof and Wink}]{collij2016application}
\bibinfo{author}{Collij, L.E.}, \bibinfo{author}{Heeman, F.},
  \bibinfo{author}{Kuijer, J.P.}, \bibinfo{author}{Ossenkoppele, R.},
  \bibinfo{author}{Benedictus, M.R.}, \bibinfo{author}{M{\"o}ller, C.},
  \bibinfo{author}{Verfaillie, S.C.}, \bibinfo{author}{Sanz-Arigita, E.J.},
  \bibinfo{author}{van Berckel, B.N.}, \bibinfo{author}{van~der Flier, W.M.}, ,
  \bibinfo{author}{Scheltens, P.}, \bibinfo{author}{Barkhof, F.},
  \bibinfo{author}{Wink, A.M.}, \bibinfo{year}{2016}.
\newblock \bibinfo{title}{Application of machine learning to arterial spin
  labeling in mild cognitive impairment and {A}lzheimer disease}.
\newblock \bibinfo{journal}{Radiology} \bibinfo{volume}{281},
  \bibinfo{pages}{865--875}.
\bibitem[{Colliot et~al.(2008)Colliot, Ch{\'e}telat, Chupin, Desgranges,
  Magnin, Benali, Dubois, Garnero, Eustache and
  Leh{\'e}ricy}]{colliot2008discrimination}
\bibinfo{author}{Colliot, O.}, \bibinfo{author}{Ch{\'e}telat, G.},
  \bibinfo{author}{Chupin, M.}, \bibinfo{author}{Desgranges, B.},
  \bibinfo{author}{Magnin, B.}, \bibinfo{author}{Benali, H.},
  \bibinfo{author}{Dubois, B.}, \bibinfo{author}{Garnero, L.},
  \bibinfo{author}{Eustache, F.}, \bibinfo{author}{Leh{\'e}ricy, S.},
  \bibinfo{year}{2008}.
\newblock \bibinfo{title}{Discrimination between alzheimer disease, mild
  cognitive impairment, and normal aging by using automated segmentation of the
  hippocampus}.
\newblock \bibinfo{journal}{Radiology} \bibinfo{volume}{248},
  \bibinfo{pages}{194--201}.
\bibitem[{Combettes and Wajs(2005)}]{Combettes2005-MMS}
\bibinfo{author}{Combettes, P.L.}, \bibinfo{author}{Wajs, V.R.},
  \bibinfo{year}{2005}.
\newblock \bibinfo{title}{{Signal Recovery by Proximal Forward-Backward
  Splitting}}.
\newblock \bibinfo{journal}{Multiscale Modeling \& Simulation}
  \bibinfo{volume}{4}, \bibinfo{pages}{1168--1200}.
\bibitem[{Costafreda et~al.(2011)Costafreda, Dinov, Tu, Shi, Liu, Kloszewska,
  Mecocci, Soininen, Tsolaki, Vellas, Wahlund, Spenger, Toga, Lovestone and
  Simmons}]{pmid21272654}
\bibinfo{author}{Costafreda, S.G.}, \bibinfo{author}{Dinov, I.D.},
  \bibinfo{author}{Tu, Z.}, \bibinfo{author}{Shi, Y.}, \bibinfo{author}{Liu,
  C.Y.}, \bibinfo{author}{Kloszewska, I.}, \bibinfo{author}{Mecocci, P.},
  \bibinfo{author}{Soininen, H.}, \bibinfo{author}{Tsolaki, M.},
  \bibinfo{author}{Vellas, B.}, \bibinfo{author}{Wahlund, L.O.},
  \bibinfo{author}{Spenger, C.}, \bibinfo{author}{Toga, A.W.},
  \bibinfo{author}{Lovestone, S.}, \bibinfo{author}{Simmons, A.},
  \bibinfo{year}{2011}.
\newblock \bibinfo{title}{{{A}utomated hippocampal shape analysis predicts the
  onset of dementia in {M}ild {C}ognitive {I}mpairment}}.
\newblock \bibinfo{journal}{Neuroimage} \bibinfo{volume}{56},
  \bibinfo{pages}{212--219}.
\bibitem[{Cuingnet et~al.(2011)Cuingnet, Gerardin, Tessieras, Auzias,
  Leh{\'e}ricy, Habert, Chupin, Benali and Colliot}]{cuingnet2011automatic}
\bibinfo{author}{Cuingnet, R.}, \bibinfo{author}{Gerardin, E.},
  \bibinfo{author}{Tessieras, J.}, \bibinfo{author}{Auzias, G.},
  \bibinfo{author}{Leh{\'e}ricy, S.}, \bibinfo{author}{Habert, M.O.},
  \bibinfo{author}{Chupin, M.}, \bibinfo{author}{Benali, H.},
  \bibinfo{author}{Colliot, O.}, \bibinfo{year}{2011}.
\newblock \bibinfo{title}{Automatic classification of patients with
  {A}lzheimer's disease from structural {MRI}: a comparison of ten methods
  using the {ADNI} database}.
\newblock \bibinfo{journal}{neuroimage} \bibinfo{volume}{56},
  \bibinfo{pages}{766--781}.
\bibitem[{De~Rotrou et~al.(2005)De~Rotrou, Wenisch, Chausson, Dray, Faucounau
  and Rigaud}]{de2005accidental}
\bibinfo{author}{De~Rotrou, J.}, \bibinfo{author}{Wenisch, E.},
  \bibinfo{author}{Chausson, C.}, \bibinfo{author}{Dray, F.},
  \bibinfo{author}{Faucounau, V.}, \bibinfo{author}{Rigaud, A.S.},
  \bibinfo{year}{2005}.
\newblock \bibinfo{title}{Accidental {MCI} in healthy subjects: a prospective
  longitudinal study}.
\newblock \bibinfo{journal}{European Journal of Neurology}
  \bibinfo{volume}{12}, \bibinfo{pages}{879--885}.
\bibitem[{Dong et~al.(2020a)Dong, Zhang, Li, Wang, Lepore, Thompson, Caselli,
  Ye and Wang}]{dongintegrating}
\bibinfo{author}{Dong, Q.}, \bibinfo{author}{Zhang, J.}, \bibinfo{author}{Li,
  Q.}, \bibinfo{author}{Wang, J.}, \bibinfo{author}{Lepore, N.},
  \bibinfo{author}{Thompson, P.M.}, \bibinfo{author}{Caselli, R.J.},
  \bibinfo{author}{Ye, J.}, \bibinfo{author}{Wang, Y.}, \bibinfo{year}{2020}a.
\newblock \bibinfo{title}{{I}ntegrating convolutional neural networks and
  multi-task dictionary learning for cognitive decline prediction with
  longitudinal images}.
\newblock \bibinfo{journal}{J. Alzheimers Dis.} \bibinfo{volume}{75},
  \bibinfo{pages}{971--992}.
\bibitem[{Dong et~al.(2020b)Dong, Zhang, Stonnington, Wu, Gutman, Chen, Su,
  Baxter, Thompson, Reiman, Caselli and Wang}]{dong2020applying}
\bibinfo{author}{Dong, Q.}, \bibinfo{author}{Zhang, W.},
  \bibinfo{author}{Stonnington, C.M.}, \bibinfo{author}{Wu, J.},
  \bibinfo{author}{Gutman, B.A.}, \bibinfo{author}{Chen, K.},
  \bibinfo{author}{Su, Y.}, \bibinfo{author}{Baxter, L.C.},
  \bibinfo{author}{Thompson, P.M.}, \bibinfo{author}{Reiman, E.M.},
  \bibinfo{author}{Caselli, R.J.}, \bibinfo{author}{Wang, Y.},
  \bibinfo{year}{2020}b.
\newblock \bibinfo{title}{{{A}pplying surface-based morphometry to study
  ventricular abnormalities of cognitively unimpaired subjects prior to
  clinically significant memory decline}}.
\newblock \bibinfo{journal}{Neuroimage Clin} \bibinfo{volume}{27},
  \bibinfo{pages}{102338}.
\bibitem[{Dong et~al.(2019)Dong, Zhang, Wu, Li, Schron, McMahon, Shi, Gutman,
  Chen, Baxter, Thompson, Reiman, Caselli and Wang}]{dong2019applying}
\bibinfo{author}{Dong, Q.}, \bibinfo{author}{Zhang, W.}, \bibinfo{author}{Wu,
  J.}, \bibinfo{author}{Li, B.}, \bibinfo{author}{Schron, E.H.},
  \bibinfo{author}{McMahon, T.}, \bibinfo{author}{Shi, J.},
  \bibinfo{author}{Gutman, B.A.}, \bibinfo{author}{Chen, K.},
  \bibinfo{author}{Baxter, L.C.}, \bibinfo{author}{Thompson, P.M.},
  \bibinfo{author}{Reiman, E.M.}, \bibinfo{author}{Caselli, R.J.},
  \bibinfo{author}{Wang, Y.}, \bibinfo{year}{2019}.
\newblock \bibinfo{title}{{{A}pplying surface-based hippocampal morphometry to
  study {A}{P}{O}{E}-{E}4 allele dose effects in cognitively unimpaired
  subjects}}.
\newblock \bibinfo{journal}{Neuroimage Clin} \bibinfo{volume}{22},
  \bibinfo{pages}{101744}.
\bibitem[{Donoho(2006)}]{donoho2006}
\bibinfo{author}{Donoho, D.L.}, \bibinfo{year}{2006}.
\newblock \bibinfo{title}{Compressed sensing}.
\newblock \bibinfo{journal}{IEEE Transactions on information theory}
  \bibinfo{volume}{52}, \bibinfo{pages}{1289--1306}.
\bibitem[{Donoho and Elad(2003)}]{pmid16576749}
\bibinfo{author}{Donoho, D.L.}, \bibinfo{author}{Elad, M.},
  \bibinfo{year}{2003}.
\newblock \bibinfo{title}{{{O}ptimally sparse representation in general
  (nonorthogonal) dictionaries via $\ell_1$ minimization}}.
\newblock \bibinfo{journal}{Proc. Natl. Acad. Sci. U.S.A.}
  \bibinfo{volume}{100}, \bibinfo{pages}{2197--2202}.
\bibitem[{Edmonds et~al.(2019)Edmonds, McDonald, Marshall, Thomas, Eppig,
  Weigand, Delano-Wood, Galasko, Salmon and Bondi}]{edmonds2019early}
\bibinfo{author}{Edmonds, E.C.}, \bibinfo{author}{McDonald, C.R.},
  \bibinfo{author}{Marshall, A.}, \bibinfo{author}{Thomas, K.R.},
  \bibinfo{author}{Eppig, J.}, \bibinfo{author}{Weigand, A.J.},
  \bibinfo{author}{Delano-Wood, L.}, \bibinfo{author}{Galasko, D.R.},
  \bibinfo{author}{Salmon, D.P.}, \bibinfo{author}{Bondi, M.W.},
  \bibinfo{year}{2019}.
\newblock \bibinfo{title}{Early versus late {MCI}: Improved {MCI} staging using
  a neuropsychological approach}.
\newblock \bibinfo{journal}{Alzheimer's \& Dementia} \bibinfo{volume}{15},
  \bibinfo{pages}{699--708}.
\bibitem[{Fan et~al.(2005)Fan, Shen and Davatzikos}]{fan:miccai05}
\bibinfo{author}{Fan, Y.}, \bibinfo{author}{Shen, D.},
  \bibinfo{author}{Davatzikos, C.}, \bibinfo{year}{2005}.
\newblock \bibinfo{title}{{{C}lassification of structural images via
  high-dimensional image warping, robust feature extraction, and {S}{V}{M}}}.
\newblock \bibinfo{journal}{Med Image Comput Comput Assist Interv}
  \bibinfo{volume}{8}, \bibinfo{pages}{1--8}.
\bibitem[{Fan et~al.(2007)Fan, Shen, Gur, Gur and Davatzikos}]{fan2007compare}
\bibinfo{author}{Fan, Y.}, \bibinfo{author}{Shen, D.}, \bibinfo{author}{Gur,
  R.C.}, \bibinfo{author}{Gur, R.E.}, \bibinfo{author}{Davatzikos, C.},
  \bibinfo{year}{2007}.
\newblock \bibinfo{title}{{{C}{O}{M}{P}{A}{R}{E}: classification of
  morphological patterns using adaptive regional elements}}.
\newblock \bibinfo{journal}{IEEE Trans Med Imaging} \bibinfo{volume}{26},
  \bibinfo{pages}{93--105}.
\bibitem[{Fan et~al.(2018)Fan, Wang, Lepore and Wang}]{fan2018tetrahedron}
\bibinfo{author}{Fan, Y.}, \bibinfo{author}{Wang, G.}, \bibinfo{author}{Lepore,
  N.}, \bibinfo{author}{Wang, Y.}, \bibinfo{year}{2018}.
\newblock \bibinfo{title}{A tetrahedron-based heat flux signature for cortical
  thickness morphometry analysis}, in: \bibinfo{booktitle}{International
  Conference on Medical Image Computing and Computer-Assisted Intervention},
  \bibinfo{organization}{Springer}. pp. \bibinfo{pages}{420--428}.
\bibitem[{Ferrarini et~al.(2008)Ferrarini, Palm, Olofsen, van~der Landen, van
  Buchem, Reiber and Admiraal-Behloul}]{pmid18228600}
\bibinfo{author}{Ferrarini, L.}, \bibinfo{author}{Palm, W.M.},
  \bibinfo{author}{Olofsen, H.}, \bibinfo{author}{van~der Landen, R.},
  \bibinfo{author}{van Buchem, M.A.}, \bibinfo{author}{Reiber, J.H.},
  \bibinfo{author}{Admiraal-Behloul, F.}, \bibinfo{year}{2008}.
\newblock \bibinfo{title}{{{V}entricular shape biomarkers for {A}lzheimer's
  disease in clinical {M}{R} images}}.
\newblock \bibinfo{journal}{Magn Reson Med} \bibinfo{volume}{59},
  \bibinfo{pages}{260--267}.
\bibitem[{Fischl(2012)}]{fischl2012freesurfer}
\bibinfo{author}{Fischl, B.}, \bibinfo{year}{2012}.
\newblock \bibinfo{title}{Freesurfer}.
\newblock \bibinfo{journal}{Neuroimage} \bibinfo{volume}{62},
  \bibinfo{pages}{774--781}.
\bibitem[{Folstein et~al.(1975)Folstein, Folstein and
  McHugh}]{folstein1975mini}
\bibinfo{author}{Folstein, M.F.}, \bibinfo{author}{Folstein, S.E.},
  \bibinfo{author}{McHugh, P.R.}, \bibinfo{year}{1975}.
\newblock \bibinfo{title}{``mini-mental state'': a practical method for grading
  the cognitive state of patients for the clinician}.
\newblock \bibinfo{journal}{Journal of psychiatric research}
  \bibinfo{volume}{12}, \bibinfo{pages}{189--198}.
\bibitem[{Freund and Schapire(1997)}]{freund1997decision}
\bibinfo{author}{Freund, Y.}, \bibinfo{author}{Schapire, R.E.},
  \bibinfo{year}{1997}.
\newblock \bibinfo{title}{A decision-theoretic generalization of on-line
  learning and an application to boosting}.
\newblock \bibinfo{journal}{Journal of computer and system sciences}
  \bibinfo{volume}{55}, \bibinfo{pages}{119--139}.
\bibitem[{Frisoni et~al.(2010)Frisoni, Fox, Jack, Scheltens and
  Thompson}]{pmid20139996}
\bibinfo{author}{Frisoni, G.B.}, \bibinfo{author}{Fox, N.C.},
  \bibinfo{author}{Jack, C.R.}, \bibinfo{author}{Scheltens, P.},
  \bibinfo{author}{Thompson, P.M.}, \bibinfo{year}{2010}.
\newblock \bibinfo{title}{{{T}he clinical use of structural {M}{R}{I} in
  {A}lzheimer disease}}.
\newblock \bibinfo{journal}{Nat Rev Neurol} \bibinfo{volume}{6},
  \bibinfo{pages}{67--77}.
\bibitem[{Gerardin et~al.(2009)Gerardin, Ch{\'e}telat, Chupin, Cuingnet,
  Desgranges, Kim, Niethammer, Dubois, Leh{\'e}ricy, Garnero, Eustache and
  Colliot}]{gerardin2009multidimensional}
\bibinfo{author}{Gerardin, E.}, \bibinfo{author}{Ch{\'e}telat, G.},
  \bibinfo{author}{Chupin, M.}, \bibinfo{author}{Cuingnet, R.},
  \bibinfo{author}{Desgranges, B.}, \bibinfo{author}{Kim, H.S.},
  \bibinfo{author}{Niethammer, M.}, \bibinfo{author}{Dubois, B.},
  \bibinfo{author}{Leh{\'e}ricy, S.}, \bibinfo{author}{Garnero, L.},
  \bibinfo{author}{Eustache, F.}, \bibinfo{author}{Colliot, O.},
  \bibinfo{year}{2009}.
\newblock \bibinfo{title}{Multidimensional classification of hippocampal shape
  features discriminates {A}lzheimer's disease and mild cognitive impairment
  from normal aging}.
\newblock \bibinfo{journal}{Neuroimage} \bibinfo{volume}{47},
  \bibinfo{pages}{1476--1486}.
\bibitem[{Guyon et~al.(2008)Guyon, Gunn, Nikravesh and
  Zadeh}]{guyon2008feature}
\bibinfo{author}{Guyon, I.}, \bibinfo{author}{Gunn, S.},
  \bibinfo{author}{Nikravesh, M.}, \bibinfo{author}{Zadeh, L.A.},
  \bibinfo{year}{2008}.
\newblock \bibinfo{title}{Feature extraction: foundations and applications}.
  volume \bibinfo{volume}{207}.
\newblock \bibinfo{publisher}{Springer}.
\bibitem[{Han et~al.(2009)Han, Xu and Prince}]{han2009a}
\bibinfo{author}{Han, X.}, \bibinfo{author}{Xu, C.}, \bibinfo{author}{Prince,
  J.L.}, \bibinfo{year}{2009}.
\newblock \bibinfo{title}{{{A} {M}oving {G}rid {F}ramework for {G}eometric
  {D}eformable {M}odels}}.
\newblock \bibinfo{journal}{Int J Comput Vis} \bibinfo{volume}{84},
  \bibinfo{pages}{63--79}.
\bibitem[{den Heijer et~al.(2010)den Heijer, van~der Lijn, Koudstaal, Hofman,
  van~der Lugt, Krestin, Niessen and Breteler}]{pmid20375138}
\bibinfo{author}{den Heijer, T.}, \bibinfo{author}{van~der Lijn, F.},
  \bibinfo{author}{Koudstaal, P.J.}, \bibinfo{author}{Hofman, A.},
  \bibinfo{author}{van~der Lugt, A.}, \bibinfo{author}{Krestin, G.P.},
  \bibinfo{author}{Niessen, W.J.}, \bibinfo{author}{Breteler, M.M.},
  \bibinfo{year}{2010}.
\newblock \bibinfo{title}{{{A} 10-year follow-up of hippocampal volume on
  magnetic resonance imaging in early dementia and cognitive decline}}.
\newblock \bibinfo{journal}{Brain} \bibinfo{volume}{133},
  \bibinfo{pages}{1163--1172}.
\bibitem[{Jack et~al.(1999)Jack, Petersen, Xu, O’Brien, Smith, Ivnik, Boeve,
  Waring, Tangalos and Kokmen}]{jack1999prediction}
\bibinfo{author}{Jack, C.}, \bibinfo{author}{Petersen, R.C.},
  \bibinfo{author}{Xu, Y.C.}, \bibinfo{author}{O’Brien, P.C.},
  \bibinfo{author}{Smith, G.E.}, \bibinfo{author}{Ivnik, R.J.},
  \bibinfo{author}{Boeve, B.F.}, \bibinfo{author}{Waring, S.C.},
  \bibinfo{author}{Tangalos, E.G.}, \bibinfo{author}{Kokmen, E.},
  \bibinfo{year}{1999}.
\newblock \bibinfo{title}{{Prediction of AD with MRI-based hippocampal volume
  in mild cognitive impairment}}.
\newblock \bibinfo{journal}{Neurology} \bibinfo{volume}{52},
  \bibinfo{pages}{1397--1397}.
\bibitem[{Jack et~al.(2003)Jack, Slomkowski, Gracon, Hoover, Felmlee, Stewart,
  Xu, Shiung, O'Brien, Cha, Knopman and Petersen}]{pmid12552040}
\bibinfo{author}{Jack, C.}, \bibinfo{author}{Slomkowski, M.},
  \bibinfo{author}{Gracon, S.}, \bibinfo{author}{Hoover, T.M.},
  \bibinfo{author}{Felmlee, J.P.}, \bibinfo{author}{Stewart, K.},
  \bibinfo{author}{Xu, Y.}, \bibinfo{author}{Shiung, M.},
  \bibinfo{author}{O'Brien, P.C.}, \bibinfo{author}{Cha, R.},
  \bibinfo{author}{Knopman, D.}, \bibinfo{author}{Petersen, R.C.},
  \bibinfo{year}{2003}.
\newblock \bibinfo{title}{{{M}{R}{I} as a biomarker of disease progression in a
  therapeutic trial of milameline for {A}{D}}}.
\newblock \bibinfo{journal}{Neurology} \bibinfo{volume}{60},
  \bibinfo{pages}{253--260}.
\bibitem[{Jack et~al.(2016)Jack, Bennett, Blennow, Carrillo, Feldman, Frisoni,
  Hampel, Jagust, Johnson, Knopman, Petersen, Scheltens, Sperling and
  Dubois}]{jack2016t}
\bibinfo{author}{Jack, C.R.}, \bibinfo{author}{Bennett, D.A.},
  \bibinfo{author}{Blennow, K.}, \bibinfo{author}{Carrillo, M.C.},
  \bibinfo{author}{Feldman, H.H.}, \bibinfo{author}{Frisoni, G.B.},
  \bibinfo{author}{Hampel, H.}, \bibinfo{author}{Jagust, W.J.},
  \bibinfo{author}{Johnson, K.A.}, \bibinfo{author}{Knopman, D.S.},
  \bibinfo{author}{Petersen, R.C.}, \bibinfo{author}{Scheltens, P.},
  \bibinfo{author}{Sperling, R.A.}, \bibinfo{author}{Dubois, B.},
  \bibinfo{year}{2016}.
\newblock \bibinfo{title}{{A/T/N: an unbiased descriptive classification scheme
  for Alzheimer disease biomarkers}}.
\newblock \bibinfo{journal}{Neurology} \bibinfo{volume}{87},
  \bibinfo{pages}{539--547}.
\bibitem[{Jain and Zongker(1997)}]{jain1997feature}
\bibinfo{author}{Jain, A.}, \bibinfo{author}{Zongker, D.},
  \bibinfo{year}{1997}.
\newblock \bibinfo{title}{Feature selection: Evaluation, application, and small
  sample performance}.
\newblock \bibinfo{journal}{IEEE transactions on pattern analysis and machine
  intelligence} \bibinfo{volume}{19}, \bibinfo{pages}{153--158}.
\bibitem[{Jiang et~al.(2015)Jiang, Li, Lv, Zhang, Zhang, Guo and
  Liu}]{pmid26466353}
\bibinfo{author}{Jiang, X.}, \bibinfo{author}{Li, X.}, \bibinfo{author}{Lv,
  J.}, \bibinfo{author}{Zhang, T.}, \bibinfo{author}{Zhang, S.},
  \bibinfo{author}{Guo, L.}, \bibinfo{author}{Liu, T.}, \bibinfo{year}{2015}.
\newblock \bibinfo{title}{{{S}parse representation of {H}{C}{P} grayordinate
  data reveals novel functional architecture of cerebral cortex}}.
\newblock \bibinfo{journal}{Hum Brain Mapp} \bibinfo{volume}{36},
  \bibinfo{pages}{5301--5319}.
\bibitem[{Jolliffe(2011)}]{jolliffe2011principal}
\bibinfo{author}{Jolliffe, I.}, \bibinfo{year}{2011}.
\newblock \bibinfo{title}{Principal component analysis}, in:
  \bibinfo{booktitle}{International encyclopedia of statistical science}.
  \bibinfo{publisher}{Springer}, pp. \bibinfo{pages}{1094--1096}.
\bibitem[{Kl{\"o}ppel et~al.(2008)Kl{\"o}ppel, Stonnington, Chu, Draganski,
  Scahill, Rohrer, Fox, Jack~Jr, Ashburner and
  Frackowiak}]{kloppel2008automatic}
\bibinfo{author}{Kl{\"o}ppel, S.}, \bibinfo{author}{Stonnington, C.M.},
  \bibinfo{author}{Chu, C.}, \bibinfo{author}{Draganski, B.},
  \bibinfo{author}{Scahill, R.I.}, \bibinfo{author}{Rohrer, J.D.},
  \bibinfo{author}{Fox, N.C.}, \bibinfo{author}{Jack~Jr, C.R.},
  \bibinfo{author}{Ashburner, J.}, \bibinfo{author}{Frackowiak, R.S.},
  \bibinfo{year}{2008}.
\newblock \bibinfo{title}{Automatic classification of {MR} scans in
  {A}lzheimer's disease}.
\newblock \bibinfo{journal}{Brain} \bibinfo{volume}{131},
  \bibinfo{pages}{681--689}.
\bibitem[{Lee et~al.(2006)Lee, Battle, Raina and Ng}]{lee2006efficient}
\bibinfo{author}{Lee, H.}, \bibinfo{author}{Battle, A.},
  \bibinfo{author}{Raina, R.}, \bibinfo{author}{Ng, A.}, \bibinfo{year}{2006}.
\newblock \bibinfo{title}{Efficient sparse coding algorithms}, in:
  \bibinfo{booktitle}{Advances in neural information processing systems}, pp.
  \bibinfo{pages}{801--808}.
\bibitem[{Leung et~al.(2010)Leung, Barnes, Ridgway, Bartlett, Clarkson,
  Macdonald, Schuff, Fox and Ourselin}]{leung2010automated}
\bibinfo{author}{Leung, K.K.}, \bibinfo{author}{Barnes, J.},
  \bibinfo{author}{Ridgway, G.R.}, \bibinfo{author}{Bartlett, J.W.},
  \bibinfo{author}{Clarkson, M.J.}, \bibinfo{author}{Macdonald, K.},
  \bibinfo{author}{Schuff, N.}, \bibinfo{author}{Fox, N.C.},
  \bibinfo{author}{Ourselin, S.}, \bibinfo{year}{2010}.
\newblock \bibinfo{title}{Automated cross-sectional and longitudinal
  hippocampal volume measurement in mild cognitive impairment and {A}lzheimer's
  disease}.
\newblock \bibinfo{journal}{NeuroImage} \bibinfo{volume}{51},
  \bibinfo{pages}{1345--1359}.
\bibitem[{Li et~al.(2014)Li, Yuan, Pu, Li, Fan, Wu, Chao, Chen, He and
  Han}]{li2014abnormal}
\bibinfo{author}{Li, S.}, \bibinfo{author}{Yuan, X.}, \bibinfo{author}{Pu, F.},
  \bibinfo{author}{Li, D.}, \bibinfo{author}{Fan, Y.}, \bibinfo{author}{Wu,
  L.}, \bibinfo{author}{Chao, W.}, \bibinfo{author}{Chen, N.},
  \bibinfo{author}{He, Y.}, \bibinfo{author}{Han, Y.}, \bibinfo{year}{2014}.
\newblock \bibinfo{title}{Abnormal changes of multidimensional surface features
  using multivariate pattern classification in amnestic mild cognitive
  impairment patients}.
\newblock \bibinfo{journal}{Journal of Neuroscience} \bibinfo{volume}{34},
  \bibinfo{pages}{10541--10553}.
\bibitem[{Li et~al.(2017)Li, Chen, Jiang, Li, Lv, Li, Peng, Tsien and
  Liu}]{li2017transcriptome}
\bibinfo{author}{Li, Y.}, \bibinfo{author}{Chen, H.}, \bibinfo{author}{Jiang,
  X.}, \bibinfo{author}{Li, X.}, \bibinfo{author}{Lv, J.}, \bibinfo{author}{Li,
  M.}, \bibinfo{author}{Peng, H.}, \bibinfo{author}{Tsien, J.Z.},
  \bibinfo{author}{Liu, T.}, \bibinfo{year}{2017}.
\newblock \bibinfo{title}{{{T}ranscriptome {A}rchitecture of {A}dult {M}ouse
  {B}rain {R}evealed by {S}parse {C}oding of {G}enome-{W}ide {I}n {S}itu
  {H}ybridization {I}mages}}.
\newblock \bibinfo{journal}{Neuroinformatics} \bibinfo{volume}{15},
  \bibinfo{pages}{285--295}.
\bibitem[{Lin et~al.(2014)Lin, Li, Sun, Lai, Davidson, Fan and
  Ye}]{lin2014stochastic}
\bibinfo{author}{Lin, B.}, \bibinfo{author}{Li, Q.}, \bibinfo{author}{Sun, Q.},
  \bibinfo{author}{Lai, M.J.}, \bibinfo{author}{Davidson, I.},
  \bibinfo{author}{Fan, W.}, \bibinfo{author}{Ye, J.}, \bibinfo{year}{2014}.
\newblock \bibinfo{title}{Stochastic coordinate coding and its application for
  drosophila gene expression pattern annotation}.
\newblock \bibinfo{journal}{arXiv preprint arXiv:1407.8147} .
\bibitem[{Liu et~al.(2011)Liu, Paajanen, Zhang, Westman, Wahlund, Simmons,
  Tunnard, Sobow, Mecocci, Tsolaki, Vellas, Muehlboeck, Evans, Spenger,
  Lovestone and Soininen}]{liu2011combination}
\bibinfo{author}{Liu, Y.}, \bibinfo{author}{Paajanen, T.},
  \bibinfo{author}{Zhang, Y.}, \bibinfo{author}{Westman, E.},
  \bibinfo{author}{Wahlund, L.O.}, \bibinfo{author}{Simmons, A.},
  \bibinfo{author}{Tunnard, C.}, \bibinfo{author}{Sobow, T.},
  \bibinfo{author}{Mecocci, P.}, \bibinfo{author}{Tsolaki, M.},
  \bibinfo{author}{Vellas, B.}, \bibinfo{author}{Muehlboeck, S.},
  \bibinfo{author}{Evans, A.}, \bibinfo{author}{Spenger, C.},
  \bibinfo{author}{Lovestone, S.}, \bibinfo{author}{Soininen, H.},
  \bibinfo{year}{2011}.
\newblock \bibinfo{title}{{{C}ombination analysis of neuropsychological tests
  and structural {M}{R}{I} measures in differentiating {A}{D}, {M}{C}{I} and
  control groups--the {A}dd{N}euro{M}ed study}}.
\newblock \bibinfo{journal}{Neurobiol. Aging} \bibinfo{volume}{32},
  \bibinfo{pages}{1198--1206}.
\bibitem[{Lorensen and Cline(1987)}]{lorensen1987marching}
\bibinfo{author}{Lorensen, W.E.}, \bibinfo{author}{Cline, H.E.},
  \bibinfo{year}{1987}.
\newblock \bibinfo{title}{Marching cubes: A high resolution {3D} surface
  construction algorithm}, in: \bibinfo{booktitle}{ACM siggraph computer
  graphics}, \bibinfo{organization}{ACM}. pp. \bibinfo{pages}{163--169}.
\bibitem[{Lv et~al.(2015a)Lv, Jiang, Li, Zhu, Chen, Zhang, Zhang, Hu, Han,
  Huang, Zhang, Guo and Liu}]{LvSparse}
\bibinfo{author}{Lv, J.}, \bibinfo{author}{Jiang, X.}, \bibinfo{author}{Li,
  X.}, \bibinfo{author}{Zhu, D.}, \bibinfo{author}{Chen, H.},
  \bibinfo{author}{Zhang, T.}, \bibinfo{author}{Zhang, S.},
  \bibinfo{author}{Hu, X.}, \bibinfo{author}{Han, J.}, \bibinfo{author}{Huang,
  H.}, \bibinfo{author}{Zhang, J.}, \bibinfo{author}{Guo, L.},
  \bibinfo{author}{Liu, T.}, \bibinfo{year}{2015}a.
\newblock \bibinfo{title}{{{S}parse representation of whole-brain f{M}{R}{I}
  signals for identification of functional networks}}.
\newblock \bibinfo{journal}{Med Image Anal} \bibinfo{volume}{20},
  \bibinfo{pages}{112--134}.
\bibitem[{Lv et~al.(2015b)Lv, Jiang, Li, Zhu, Zhang, Zhao, Chen, Zhang, Hu,
  Han, Ye, Guo and Liu}]{lv2015holistic}
\bibinfo{author}{Lv, J.}, \bibinfo{author}{Jiang, X.}, \bibinfo{author}{Li,
  X.}, \bibinfo{author}{Zhu, D.}, \bibinfo{author}{Zhang, S.},
  \bibinfo{author}{Zhao, S.}, \bibinfo{author}{Chen, H.},
  \bibinfo{author}{Zhang, T.}, \bibinfo{author}{Hu, X.}, \bibinfo{author}{Han,
  J.}, \bibinfo{author}{Ye, J.}, \bibinfo{author}{Guo, L.},
  \bibinfo{author}{Liu, T.}, \bibinfo{year}{2015}b.
\newblock \bibinfo{title}{{{H}olistic atlases of functional networks and
  interactions reveal reciprocal organizational architecture of cortical
  function}}.
\newblock \bibinfo{journal}{IEEE Trans Biomed Eng} \bibinfo{volume}{62},
  \bibinfo{pages}{1120--1131}.
\bibitem[{Lv et~al.(2017)Lv, Lin, Li, Zhang, Zhao, Jiang, Guo, Han, Hu, Guo, Ye
  and Liu}]{Lvtask}
\bibinfo{author}{Lv, J.}, \bibinfo{author}{Lin, B.}, \bibinfo{author}{Li, Q.},
  \bibinfo{author}{Zhang, W.}, \bibinfo{author}{Zhao, Y.},
  \bibinfo{author}{Jiang, X.}, \bibinfo{author}{Guo, L.}, \bibinfo{author}{Han,
  J.}, \bibinfo{author}{Hu, X.}, \bibinfo{author}{Guo, C.},
  \bibinfo{author}{Ye, J.}, \bibinfo{author}{Liu, T.}, \bibinfo{year}{2017}.
\newblock \bibinfo{title}{{{T}ask f{M}{R}{I} data analysis based on supervised
  stochastic coordinate coding}}.
\newblock \bibinfo{journal}{Med Image Anal} \bibinfo{volume}{38},
  \bibinfo{pages}{1--16}.
\bibitem[{Magnin et~al.(2009)Magnin, Mesrob, Kinkingn{\'e}hun,
  P{\'e}l{\'e}grini-Issac, Colliot, Sarazin, Dubois, Leh{\'e}ricy and
  Benali}]{magnin2009support}
\bibinfo{author}{Magnin, B.}, \bibinfo{author}{Mesrob, L.},
  \bibinfo{author}{Kinkingn{\'e}hun, S.},
  \bibinfo{author}{P{\'e}l{\'e}grini-Issac, M.}, \bibinfo{author}{Colliot, O.},
  \bibinfo{author}{Sarazin, M.}, \bibinfo{author}{Dubois, B.},
  \bibinfo{author}{Leh{\'e}ricy, S.}, \bibinfo{author}{Benali, H.},
  \bibinfo{year}{2009}.
\newblock \bibinfo{title}{Support vector machine-based classification of
  {A}lzheimer’s disease from whole-brain anatomical {MRI}}.
\newblock \bibinfo{journal}{Neuroradiology} \bibinfo{volume}{51},
  \bibinfo{pages}{73--83}.
\bibitem[{Mairal et~al.(2009)Mairal, Bach, Ponce and Sapiro}]{mairal2009online}
\bibinfo{author}{Mairal, J.}, \bibinfo{author}{Bach, F.},
  \bibinfo{author}{Ponce, J.}, \bibinfo{author}{Sapiro, G.},
  \bibinfo{year}{2009}.
\newblock \bibinfo{title}{Online dictionary learning for sparse coding}, in:
  \bibinfo{booktitle}{Proceedings of the 26th annual international conference
  on machine learning}, \bibinfo{organization}{ACM}. pp.
  \bibinfo{pages}{689--696}.
\bibitem[{Mi et~al.(2020)Mi, Zhang and Wang}]{Mi:AAAI20}
\bibinfo{author}{Mi, L.}, \bibinfo{author}{Zhang, W.}, \bibinfo{author}{Wang,
  Y.}, \bibinfo{year}{2020}.
\newblock \bibinfo{title}{Regularized wasserstein means for aligning
  distributional data}, in: \bibinfo{booktitle}{Proceedings of 34th Conference
  on Artificial Intelligence (AAAI-20)}, pp. \bibinfo{pages}{5166--5173}.
\bibitem[{Moenning and Dodgson(2003)}]{moenning2003fast}
\bibinfo{author}{Moenning, C.}, \bibinfo{author}{Dodgson, N.A.},
  \bibinfo{year}{2003}.
\newblock \bibinfo{title}{{Fast Marching farthest point sampling}}.
\newblock \bibinfo{type}{Technical Report} \bibinfo{number}{UCAM-CL-TR-562}.
  University of Cambridge, Computer Laboratory.
\newblock \URLprefix
  \url{https://www.cl.cam.ac.uk/techreports/UCAM-CL-TR-562.pdf}.
\bibitem[{Moradi et~al.(2015)Moradi, Pepe, Gaser, Huttunen and
  Tohka}]{moradi2015machine}
\bibinfo{author}{Moradi, E.}, \bibinfo{author}{Pepe, A.},
  \bibinfo{author}{Gaser, C.}, \bibinfo{author}{Huttunen, H.},
  \bibinfo{author}{Tohka, J.}, \bibinfo{year}{2015}.
\newblock \bibinfo{title}{Machine learning framework for early {MRI}-based
  {A}lzheimer's conversion prediction in {MCI} subjects}.
\newblock \bibinfo{journal}{Neuroimage} \bibinfo{volume}{104},
  \bibinfo{pages}{398--412}.
\bibitem[{Patel et~al.(2015)Patel, Shah, Shingadiya and
  Patel}]{patelcomparison}
\bibinfo{author}{Patel, J.R.}, \bibinfo{author}{Shah, T.R.},
  \bibinfo{author}{Shingadiya, V.P.}, \bibinfo{author}{Patel, V.B.},
  \bibinfo{year}{2015}.
\newblock \bibinfo{title}{{{C}omparison between breadth first search and
  nearest neighbor algorithm for waveguide path planning}}.
\newblock \bibinfo{journal}{Int. J. Research and Scientific Innovation}
  \bibinfo{volume}{2}, \bibinfo{pages}{19--21}.
\bibitem[{Petersen et~al.(2010)Petersen, Aisen, Beckett, Donohue, Gamst,
  Harvey, Jack, Jagust, Shaw, Toga et~al.}]{petersen2010alzheimer}
\bibinfo{author}{Petersen, R.C.}, \bibinfo{author}{Aisen, P.},
  \bibinfo{author}{Beckett, L.A.}, \bibinfo{author}{Donohue, M.},
  \bibinfo{author}{Gamst, A.}, \bibinfo{author}{Harvey, D.J.},
  \bibinfo{author}{Jack, C.}, \bibinfo{author}{Jagust, W.},
  \bibinfo{author}{Shaw, L.}, \bibinfo{author}{Toga, A.}, et~al.,
  \bibinfo{year}{2010}.
\newblock \bibinfo{title}{Alzheimer's disease neuroimaging initiative {(ADNI)}:
  clinical characterization}.
\newblock \bibinfo{journal}{Neurology} \bibinfo{volume}{74},
  \bibinfo{pages}{201--209}.
\bibitem[{Qiu et~al.(2010)Qiu, Brown, Fischl, Ma and Miller}]{pmid20129863}
\bibinfo{author}{Qiu, A.}, \bibinfo{author}{Brown, T.},
  \bibinfo{author}{Fischl, B.}, \bibinfo{author}{Ma, J.},
  \bibinfo{author}{Miller, M.I.}, \bibinfo{year}{2010}.
\newblock \bibinfo{title}{{{A}tlas generation for subcortical and ventricular
  structures with its applications in shape analysis}}.
\newblock \bibinfo{journal}{IEEE Trans Image Process} \bibinfo{volume}{19},
  \bibinfo{pages}{1539--1547}.
\bibitem[{Rathore et~al.(2017)Rathore, Habes, Iftikhar, Shacklett and
  Davatzikos}]{rathore2017review}
\bibinfo{author}{Rathore, S.}, \bibinfo{author}{Habes, M.},
  \bibinfo{author}{Iftikhar, M.A.}, \bibinfo{author}{Shacklett, A.},
  \bibinfo{author}{Davatzikos, C.}, \bibinfo{year}{2017}.
\newblock \bibinfo{title}{A review on neuroimaging-based classification studies
  and associated feature extraction methods for alzheimer's disease and its
  prodromal stages}.
\newblock \bibinfo{journal}{NeuroImage} \bibinfo{volume}{155},
  \bibinfo{pages}{530--548}.
\bibitem[{Rey(1964)}]{Rey:1964}
\bibinfo{author}{Rey, A.}, \bibinfo{year}{1964}.
\newblock \bibinfo{title}{L’examen clinique en psychologie}.
\newblock \bibinfo{publisher}{Presses Universitaires de France; Paris}.
\bibitem[{Rojas(2009)}]{rojas2009adaboost}
\bibinfo{author}{Rojas, R.}, \bibinfo{year}{2009}.
\newblock \bibinfo{title}{Adaboost and the super bowl of classifiers a tutorial
  introduction to adaptive boosting}.
\newblock \bibinfo{journal}{Freie University, Berlin, Tech. Rep} .
\bibitem[{Rosen et~al.(1984)Rosen, Mohs and Davis}]{rosen1984new}
\bibinfo{author}{Rosen, W.G.}, \bibinfo{author}{Mohs, R.C.},
  \bibinfo{author}{Davis, K.L.}, \bibinfo{year}{1984}.
\newblock \bibinfo{title}{A new rating scale for {A}lzheimer's disease.}
\newblock \bibinfo{journal}{The American journal of psychiatry} .
\bibitem[{Saadi et~al.(2007)Saadi, Talbot and Cawley}]{saadi2007optimally}
\bibinfo{author}{Saadi, K.}, \bibinfo{author}{Talbot, N.L.},
  \bibinfo{author}{Cawley, G.C.}, \bibinfo{year}{2007}.
\newblock \bibinfo{title}{{{O}ptimally regularised kernel {F}isher discriminant
  classification}}.
\newblock \bibinfo{journal}{Neural Netw} \bibinfo{volume}{20},
  \bibinfo{pages}{832--841}.
\bibitem[{Scholkopft and Mullert(1999)}]{scholkopft1999fisher}
\bibinfo{author}{Scholkopft, B.}, \bibinfo{author}{Mullert, K.R.},
  \bibinfo{year}{1999}.
\newblock \bibinfo{title}{Fisher discriminant analysis with kernels}.
\newblock \bibinfo{journal}{Neural networks for signal processing IX}
  \bibinfo{volume}{1}, \bibinfo{pages}{1}.
\bibitem[{Shen et~al.(2012)Shen, Fripp, M{\'e}riaudeau, Ch{\'e}telat, Salvado
  and Bourgeat}]{shen2012detecting}
\bibinfo{author}{Shen, K.K.}, \bibinfo{author}{Fripp, J.},
  \bibinfo{author}{M{\'e}riaudeau, F.}, \bibinfo{author}{Ch{\'e}telat, G.},
  \bibinfo{author}{Salvado, O.}, \bibinfo{author}{Bourgeat, P.},
  \bibinfo{year}{2012}.
\newblock \bibinfo{title}{Detecting global and local hippocampal shape changes
  in {A}lzheimer's disease using statistical shape models}.
\newblock \bibinfo{journal}{Neuroimage} \bibinfo{volume}{59},
  \bibinfo{pages}{2155--2166}.
\bibitem[{Shi et~al.(2015)Shi, Stonnington, Thompson, Chen, Gutman, Reschke,
  Baxter, Reiman, Caselli and Wang}]{shi2015studying}
\bibinfo{author}{Shi, J.}, \bibinfo{author}{Stonnington, C.M.},
  \bibinfo{author}{Thompson, P.M.}, \bibinfo{author}{Chen, K.},
  \bibinfo{author}{Gutman, B.}, \bibinfo{author}{Reschke, C.},
  \bibinfo{author}{Baxter, L.C.}, \bibinfo{author}{Reiman, E.M.},
  \bibinfo{author}{Caselli, R.J.}, \bibinfo{author}{Wang, Y.},
  \bibinfo{year}{2015}.
\newblock \bibinfo{title}{Studying ventricular abnormalities in {M}ild
  {C}ognitive {I}mpairment with hyperbolic {R}icci flow and {T}ensor-based
  morphometry}.
\newblock \bibinfo{journal}{NeuroImage} \bibinfo{volume}{104},
  \bibinfo{pages}{1--20}.
\bibitem[{Shi et~al.(2013)Shi, Thompson, Gutman and Wang}]{shi2013surface}
\bibinfo{author}{Shi, J.}, \bibinfo{author}{Thompson, P.M.},
  \bibinfo{author}{Gutman, B.}, \bibinfo{author}{Wang, Y.},
  \bibinfo{year}{2013}.
\newblock \bibinfo{title}{{{S}urface fluid registration of conformal
  representation: application to detect disease burden and genetic influence on
  hippocampus}}.
\newblock \bibinfo{journal}{Neuroimage} \bibinfo{volume}{78},
  \bibinfo{pages}{111--134}.
\bibitem[{Shi and Wang(2020)}]{shi2019hyperbolic}
\bibinfo{author}{Shi, J.}, \bibinfo{author}{Wang, Y.}, \bibinfo{year}{2020}.
\newblock \bibinfo{title}{{{H}yperbolic {W}asserstein {D}istance for {S}hape
  {I}ndexing}}.
\newblock \bibinfo{journal}{IEEE Trans Pattern Anal Mach Intell}
  \bibinfo{volume}{42}, \bibinfo{pages}{1362--1376}.
\bibitem[{Shi et~al.(2017)Shi, Zhang, Tang, Caselli and
  Wang}]{shi2017conformal}
\bibinfo{author}{Shi, J.}, \bibinfo{author}{Zhang, W.}, \bibinfo{author}{Tang,
  M.}, \bibinfo{author}{Caselli, R.J.}, \bibinfo{author}{Wang, Y.},
  \bibinfo{year}{2017}.
\newblock \bibinfo{title}{{{C}onformal invariants for multiply connected
  surfaces: {A}pplication to landmark curve-based brain morphometry analysis}}.
\newblock \bibinfo{journal}{Med Image Anal} \bibinfo{volume}{35},
  \bibinfo{pages}{517--529}.
\bibitem[{S{\o}rensen et~al.(2016)S{\o}rensen, Igel, Liv~Hansen, Osler,
  Lauritzen, Rostrup and Nielsen}]{sorensen2016early}
\bibinfo{author}{S{\o}rensen, L.}, \bibinfo{author}{Igel, C.},
  \bibinfo{author}{Liv~Hansen, N.}, \bibinfo{author}{Osler, M.},
  \bibinfo{author}{Lauritzen, M.}, \bibinfo{author}{Rostrup, E.},
  \bibinfo{author}{Nielsen, M.}, \bibinfo{year}{2016}.
\newblock \bibinfo{title}{Early detection of {A}lzheimer's disease using {MRI}
  hippocampal texture}.
\newblock \bibinfo{journal}{Human Brain Papping} \bibinfo{volume}{37},
  \bibinfo{pages}{1148--1161}.
\bibitem[{Styner et~al.(2005)Styner, Lieberman, McClure, Weinberger, Jones and
  Gerig}]{pmid15772166}
\bibinfo{author}{Styner, M.}, \bibinfo{author}{Lieberman, J.A.},
  \bibinfo{author}{McClure, R.K.}, \bibinfo{author}{Weinberger, D.R.},
  \bibinfo{author}{Jones, D.W.}, \bibinfo{author}{Gerig, G.},
  \bibinfo{year}{2005}.
\newblock \bibinfo{title}{{{M}orphometric analysis of lateral ventricles in
  schizophrenia and healthy controls regarding genetic and disease-specific
  factors}}.
\newblock \bibinfo{journal}{Proc. Natl. Acad. Sci. U.S.A.}
  \bibinfo{volume}{102}, \bibinfo{pages}{4872--4877}.
\bibitem[{Sun et~al.(2009)Sun, van Erp, Thompson, Bearden, Daley, Kushan,
  Hardt, Nuechterlein, Toga and Cannon}]{sun2009elucidating}
\bibinfo{author}{Sun, D.}, \bibinfo{author}{van Erp, T.G.},
  \bibinfo{author}{Thompson, P.M.}, \bibinfo{author}{Bearden, C.E.},
  \bibinfo{author}{Daley, M.}, \bibinfo{author}{Kushan, L.},
  \bibinfo{author}{Hardt, M.E.}, \bibinfo{author}{Nuechterlein, K.H.},
  \bibinfo{author}{Toga, A.W.}, \bibinfo{author}{Cannon, T.D.},
  \bibinfo{year}{2009}.
\newblock \bibinfo{title}{{{E}lucidating a magnetic resonance imaging-based
  neuroanatomic biomarker for psychosis: classification analysis using
  probabilistic brain atlas and machine learning algorithms}}.
\newblock \bibinfo{journal}{Biol. Psychiatry} \bibinfo{volume}{66},
  \bibinfo{pages}{1055--1060}.
\bibitem[{Thompson et~al.(2000)Thompson, Giedd, Woods, MacDonald, Evans and
  Toga}]{thompson2000growth}
\bibinfo{author}{Thompson, P.M.}, \bibinfo{author}{Giedd, J.N.},
  \bibinfo{author}{Woods, R.P.}, \bibinfo{author}{MacDonald, D.},
  \bibinfo{author}{Evans, A.C.}, \bibinfo{author}{Toga, A.W.},
  \bibinfo{year}{2000}.
\newblock \bibinfo{title}{Growth patterns in the developing brain detected by
  using continuum mechanical tensor maps}.
\newblock \bibinfo{journal}{Nature} \bibinfo{volume}{404},
  \bibinfo{pages}{190--193}.
\bibitem[{Thompson et~al.(2004a)Thompson, Hayashi, De~Zubicaray, Janke, Rose,
  Semple, Hong, Herman, Gravano, Doddrell and Toga}]{pmid15275931}
\bibinfo{author}{Thompson, P.M.}, \bibinfo{author}{Hayashi, K.M.},
  \bibinfo{author}{De~Zubicaray, G.I.}, \bibinfo{author}{Janke, A.L.},
  \bibinfo{author}{Rose, S.E.}, \bibinfo{author}{Semple, J.},
  \bibinfo{author}{Hong, M.S.}, \bibinfo{author}{Herman, D.H.},
  \bibinfo{author}{Gravano, D.}, \bibinfo{author}{Doddrell, D.M.},
  \bibinfo{author}{Toga, A.W.}, \bibinfo{year}{2004}a.
\newblock \bibinfo{title}{{{M}apping hippocampal and ventricular change in
  {A}lzheimer disease}}.
\newblock \bibinfo{journal}{Neuroimage} \bibinfo{volume}{22},
  \bibinfo{pages}{1754--1766}.
\bibitem[{Thompson et~al.(2004b)Thompson, Hayashi, Sowell, Gogtay, Giedd,
  Rapoport, de~Zubicaray, Janke, Rose, Semple, Doddrell, Wang, van Erp, Cannon
  and Toga}]{thompson2004mapping}
\bibinfo{author}{Thompson, P.M.}, \bibinfo{author}{Hayashi, K.M.},
  \bibinfo{author}{Sowell, E.R.}, \bibinfo{author}{Gogtay, N.},
  \bibinfo{author}{Giedd, J.N.}, \bibinfo{author}{Rapoport, J.L.},
  \bibinfo{author}{de~Zubicaray, G.I.}, \bibinfo{author}{Janke, A.L.},
  \bibinfo{author}{Rose, S.E.}, \bibinfo{author}{Semple, J.},
  \bibinfo{author}{Doddrell, D.M.}, \bibinfo{author}{Wang, Y.},
  \bibinfo{author}{van Erp, T.G.}, \bibinfo{author}{Cannon, T.D.},
  \bibinfo{author}{Toga, A.W.}, \bibinfo{year}{2004}b.
\newblock \bibinfo{title}{{{M}apping cortical change in {A}lzheimer's disease,
  brain development, and schizophrenia}}.
\newblock \bibinfo{journal}{Neuroimage} \bibinfo{volume}{23 Suppl 1},
  \bibinfo{pages}{2--18}.
\bibitem[{Tibshirani(1994)}]{Tibshirani94regressionshrinkage}
\bibinfo{author}{Tibshirani, R.}, \bibinfo{year}{1994}.
\newblock \bibinfo{title}{Regression shrinkage and selection via the {LASSO}}.
\newblock \bibinfo{journal}{Journal of the Royal Statistical Society, Series B}
  \bibinfo{volume}{58}, \bibinfo{pages}{267--288}.
\bibitem[{Tosun et~al.(2016)Tosun, Chen, Yu, Sundell, Suhy, Siemers, Schwarz
  and Weiner}]{tosun2016amyloid}
\bibinfo{author}{Tosun, D.}, \bibinfo{author}{Chen, Y.F.}, \bibinfo{author}{Yu,
  P.}, \bibinfo{author}{Sundell, K.L.}, \bibinfo{author}{Suhy, J.},
  \bibinfo{author}{Siemers, E.}, \bibinfo{author}{Schwarz, A.J.},
  \bibinfo{author}{Weiner, M.W.}, \bibinfo{year}{2016}.
\newblock \bibinfo{title}{Amyloid status imputed from a multimodal classifier
  including structural mri distinguishes progressors from nonprogressors in a
  mild alzheimer's disease clinical trial cohort}.
\newblock \bibinfo{journal}{Alzheimer's \& Dementia} \bibinfo{volume}{12},
  \bibinfo{pages}{977--986}.
\bibitem[{Tosun et~al.(2014)Tosun, Joshi and Weiner}]{tosun2014multimodal}
\bibinfo{author}{Tosun, D.}, \bibinfo{author}{Joshi, S.},
  \bibinfo{author}{Weiner, M.W.}, \bibinfo{year}{2014}.
\newblock \bibinfo{title}{Multimodal {MRI}-based imputation of the {A}$\beta$+
  in early mild cognitive impairment}.
\newblock \bibinfo{journal}{Annals of clinical and translational neurology}
  \bibinfo{volume}{1}, \bibinfo{pages}{160--170}.
\bibitem[{Tsui et~al.(2013)Tsui, Fenton, Vuong, Hass, Koehl, Amenta,
  Coeurjolly, DeCarli and Carmichael}]{Tsui:IPMI13}
\bibinfo{author}{Tsui, A.}, \bibinfo{author}{Fenton, D.},
  \bibinfo{author}{Vuong, P.}, \bibinfo{author}{Hass, J.},
  \bibinfo{author}{Koehl, P.}, \bibinfo{author}{Amenta, N.},
  \bibinfo{author}{Coeurjolly, D.}, \bibinfo{author}{DeCarli, C.},
  \bibinfo{author}{Carmichael, O.}, \bibinfo{year}{2013}.
\newblock \bibinfo{title}{{{G}lobally optimal cortical surface matching with
  exact landmark correspondence}}.
\newblock \bibinfo{journal}{Inf Process Med Imaging} \bibinfo{volume}{23},
  \bibinfo{pages}{487--498}.
\bibitem[{Van~Essen(2012)}]{van2012cortical}
\bibinfo{author}{Van~Essen, D.C.}, \bibinfo{year}{2012}.
\newblock \bibinfo{title}{Cortical cartography and caret software}.
\newblock \bibinfo{journal}{Neuroimage} \bibinfo{volume}{62},
  \bibinfo{pages}{757--764}.
\bibitem[{Varatharajah et~al.(2019)Varatharajah, Ramanan, Iyer and
  Vemuri}]{varatharajah2019predicting}
\bibinfo{author}{Varatharajah, Y.}, \bibinfo{author}{Ramanan, V.K.},
  \bibinfo{author}{Iyer, R.}, \bibinfo{author}{Vemuri, P.},
  \bibinfo{year}{2019}.
\newblock \bibinfo{title}{{Predicting short-term MCI-to-AD progression using
  imaging, CSF, genetic factors, cognitive resilience, and demographics}}.
\newblock \bibinfo{journal}{Scientific reports} \bibinfo{volume}{9},
  \bibinfo{pages}{1--15}.
\bibitem[{Vemuri et~al.(2008)Vemuri, Gunter, Senjem, Whitwell, Kantarci,
  Knopman, Boeve, Petersen and Jack}]{pmid18054253}
\bibinfo{author}{Vemuri, P.}, \bibinfo{author}{Gunter, J.L.},
  \bibinfo{author}{Senjem, M.L.}, \bibinfo{author}{Whitwell, J.L.},
  \bibinfo{author}{Kantarci, K.}, \bibinfo{author}{Knopman, D.S.},
  \bibinfo{author}{Boeve, B.F.}, \bibinfo{author}{Petersen, R.C.},
  \bibinfo{author}{Jack, C.R.}, \bibinfo{year}{2008}.
\newblock \bibinfo{title}{{A}lzheimer's disease diagnosis in individual
  subjects using structural {M}{R} images: validation studies}.
\newblock \bibinfo{journal}{Neuroimage} \bibinfo{volume}{39},
  \bibinfo{pages}{1186--1197}.
\bibitem[{Vounou et~al.(2010)Vounou, Nichols, Montana, Initiative
  et~al.}]{vounou2010discovering}
\bibinfo{author}{Vounou, M.}, \bibinfo{author}{Nichols, T.E.},
  \bibinfo{author}{Montana, G.}, \bibinfo{author}{Initiative, A.D.N.}, et~al.,
  \bibinfo{year}{2010}.
\newblock \bibinfo{title}{{D}iscovering genetic associations with
  high-dimensional neuroimaging phenotypes: {A} sparse reduced-rank regression
  approach}.
\newblock \bibinfo{journal}{Neuroimage} \bibinfo{volume}{53},
  \bibinfo{pages}{1147--1159}.
\bibitem[{Vu and Monga(2017)}]{vu2017fast}
\bibinfo{author}{Vu, T.H.}, \bibinfo{author}{Monga, V.}, \bibinfo{year}{2017}.
\newblock \bibinfo{title}{Fast low-rank shared dictionary learning for image
  classification}.
\newblock \bibinfo{journal}{IEEE Transactions on Image Processing}
  \bibinfo{volume}{26}, \bibinfo{pages}{5160--5175}.
\bibitem[{Wang et~al.(2011)Wang, Song, Rajagopalan, An, Liu, Chou, Gutman,
  Toga, Thompson, Initiative et~al.}]{wang2011surface}
\bibinfo{author}{Wang, Y.}, \bibinfo{author}{Song, Y.},
  \bibinfo{author}{Rajagopalan, P.}, \bibinfo{author}{An, T.},
  \bibinfo{author}{Liu, K.}, \bibinfo{author}{Chou, Y.Y.},
  \bibinfo{author}{Gutman, B.}, \bibinfo{author}{Toga, A.W.},
  \bibinfo{author}{Thompson, P.M.}, \bibinfo{author}{Initiative, A.D.N.},
  et~al., \bibinfo{year}{2011}.
\newblock \bibinfo{title}{{{S}urface-based {T}{B}{M} boosts power to detect
  disease effects on the brain: an {N}=804 {A}{D}{N}{I} study}}.
\newblock \bibinfo{journal}{Neuroimage} \bibinfo{volume}{56},
  \bibinfo{pages}{1993--2010}.
\bibitem[{Wang et~al.(2013)Wang, Yuan, Shi, Greve, Ye, Toga, Reiss and
  Thompson}]{wang2013applying}
\bibinfo{author}{Wang, Y.}, \bibinfo{author}{Yuan, L.}, \bibinfo{author}{Shi,
  J.}, \bibinfo{author}{Greve, A.}, \bibinfo{author}{Ye, J.},
  \bibinfo{author}{Toga, A.W.}, \bibinfo{author}{Reiss, A.L.},
  \bibinfo{author}{Thompson, P.M.}, \bibinfo{year}{2013}.
\newblock \bibinfo{title}{{{A}pplying tensor-based morphometry to parametric
  surfaces can improve {M}{R}{I}-based disease diagnosis}}.
\newblock \bibinfo{journal}{Neuroimage} \bibinfo{volume}{74},
  \bibinfo{pages}{209--230}.
\bibitem[{Wang et~al.(2010)Wang, Zhang, Gutman, Chan, Becker, Aizenstein,
  Lopez, Tamburo, Toga and Thompson}]{Wang:NIMG10}
\bibinfo{author}{Wang, Y.}, \bibinfo{author}{Zhang, J.},
  \bibinfo{author}{Gutman, B.}, \bibinfo{author}{Chan, T.F.},
  \bibinfo{author}{Becker, J.T.}, \bibinfo{author}{Aizenstein, H.J.},
  \bibinfo{author}{Lopez, O.L.}, \bibinfo{author}{Tamburo, R.J.},
  \bibinfo{author}{Toga, A.W.}, \bibinfo{author}{Thompson, P.M.},
  \bibinfo{year}{2010}.
\newblock \bibinfo{title}{{{M}ultivariate tensor-based morphometry on surfaces:
  application to mapping ventricular abnormalities in {H}{I}{V}/{A}{I}{D}{S}}}.
\newblock \bibinfo{journal}{Neuroimage} \bibinfo{volume}{49},
  \bibinfo{pages}{2141--2157}.
\bibitem[{Wee et~al.(2019)Wee, Liu, Lee, Poh, Ji, Qiu, Initiative
  et~al.}]{wee2019cortical}
\bibinfo{author}{Wee, C.Y.}, \bibinfo{author}{Liu, C.}, \bibinfo{author}{Lee,
  A.}, \bibinfo{author}{Poh, J.S.}, \bibinfo{author}{Ji, H.},
  \bibinfo{author}{Qiu, A.}, \bibinfo{author}{Initiative, A.D.N.}, et~al.,
  \bibinfo{year}{2019}.
\newblock \bibinfo{title}{Cortical graph neural network for ad and mci
  diagnosis and transfer learning across populations}.
\newblock \bibinfo{journal}{NeuroImage: Clinical} \bibinfo{volume}{23},
  \bibinfo{pages}{101929}.
\bibitem[{Weiner et~al.(2013)Weiner, Veitch, Aisen, Beckett, Cairns, Green,
  Harvey, Jack, Jagust, Liu, Morris, Petersen, Saykin, Schmidt, Shaw, Shen,
  Siuciak, Soares, Toga and Trojanowski}]{ADNI}
\bibinfo{author}{Weiner, M.W.}, \bibinfo{author}{Veitch, D.P.},
  \bibinfo{author}{Aisen, P.S.}, \bibinfo{author}{Beckett, L.A.},
  \bibinfo{author}{Cairns, N.J.}, \bibinfo{author}{Green, R.C.},
  \bibinfo{author}{Harvey, D.}, \bibinfo{author}{Jack, C.R.},
  \bibinfo{author}{Jagust, W.}, \bibinfo{author}{Liu, E.},
  \bibinfo{author}{Morris, J.C.}, \bibinfo{author}{Petersen, R.C.},
  \bibinfo{author}{Saykin, A.J.}, \bibinfo{author}{Schmidt, M.E.},
  \bibinfo{author}{Shaw, L.}, \bibinfo{author}{Shen, L.},
  \bibinfo{author}{Siuciak, J.A.}, \bibinfo{author}{Soares, H.},
  \bibinfo{author}{Toga, A.W.}, \bibinfo{author}{Trojanowski, J.Q.},
  \bibinfo{year}{2013}.
\newblock \bibinfo{title}{{{T}he {A}lzheimer's {D}isease {N}euroimaging
  {I}nitiative: a review of papers published since its inception}}.
\newblock \bibinfo{journal}{Alzheimers Dement} \bibinfo{volume}{9},
  \bibinfo{pages}{e111--194}.
\bibitem[{Wolz et~al.(2010)Wolz, Heckemann, Aljabar, Hajnal, Hammers, Lotjonen
  and Rueckert}]{pmid20382238}
\bibinfo{author}{Wolz, R.}, \bibinfo{author}{Heckemann, R.A.},
  \bibinfo{author}{Aljabar, P.}, \bibinfo{author}{Hajnal, J.V.},
  \bibinfo{author}{Hammers, A.}, \bibinfo{author}{Lotjonen, J.},
  \bibinfo{author}{Rueckert, D.}, \bibinfo{year}{2010}.
\newblock \bibinfo{title}{{{M}easurement of hippocampal atrophy using 4{D}
  graph-cut segmentation: application to {A}{D}{N}{I}}}.
\newblock \bibinfo{journal}{Neuroimage} \bibinfo{volume}{52},
  \bibinfo{pages}{109--118}.
\bibitem[{Wu et~al.(2018)Wu, Zhang, Shi, Chen, Caselli, Reiman and
  Wang}]{wu2018hippocampus}
\bibinfo{author}{Wu, J.}, \bibinfo{author}{Zhang, J.}, \bibinfo{author}{Shi,
  J.}, \bibinfo{author}{Chen, K.}, \bibinfo{author}{Caselli, R.J.},
  \bibinfo{author}{Reiman, E.M.}, \bibinfo{author}{Wang, Y.},
  \bibinfo{year}{2018}.
\newblock \bibinfo{title}{Hippocampus morphometry study on pathology-confirmed
  alzheimer's disease patients with surface multivariate morphometry
  statistics}, in: \bibinfo{booktitle}{2018 IEEE 15th International Symposium
  on Biomedical Imaging (ISBI 2018)}, \bibinfo{organization}{IEEE}. pp.
  \bibinfo{pages}{1555--1559}.
\bibitem[{Wu and Lange(2008)}]{TTW08a}
\bibinfo{author}{Wu, T.T.}, \bibinfo{author}{Lange, K.}, \bibinfo{year}{2008}.
\newblock \bibinfo{title}{{Coordinate Descent Algorithms for LASSO Penalized
  Regression}}.
\newblock \bibinfo{journal}{The Annals of Applied Statistics}
  \bibinfo{volume}{2}, \bibinfo{pages}{224--244}.
\bibitem[{Yang et~al.(2010)Yang, Wright, Huang and Ma}]{yang2010image}
\bibinfo{author}{Yang, J.}, \bibinfo{author}{Wright, J.},
  \bibinfo{author}{Huang, T.S.}, \bibinfo{author}{Ma, Y.},
  \bibinfo{year}{2010}.
\newblock \bibinfo{title}{{{I}mage super-resolution via sparse
  representation}}.
\newblock \bibinfo{journal}{IEEE Trans Image Process} \bibinfo{volume}{19},
  \bibinfo{pages}{2861--2873}.
\bibitem[{Yin et~al.(2008)Yin, Osher, Goldfarb and Darbon}]{yin2008}
\bibinfo{author}{Yin, W.}, \bibinfo{author}{Osher, S.},
  \bibinfo{author}{Goldfarb, D.}, \bibinfo{author}{Darbon, J.},
  \bibinfo{year}{2008}.
\newblock \bibinfo{title}{Bregman iterative algorithms for
  $\ell_1$-minimization with applications to compressed sensing}.
\newblock \bibinfo{journal}{SIAM Journal on Imaging sciences}
  \bibinfo{volume}{1}, \bibinfo{pages}{143--168}.
\bibitem[{Zeng et~al.(2013)Zeng, Shi, Wang, Yau, Gu and {Alzheimer's Disease
  Neuroimaging Initiative}}]{Zeng2013}
\bibinfo{author}{Zeng, W.}, \bibinfo{author}{Shi, R.}, \bibinfo{author}{Wang,
  Y.}, \bibinfo{author}{Yau, S.T.}, \bibinfo{author}{Gu, X.},
  \bibinfo{author}{{Alzheimer's Disease Neuroimaging Initiative}},
  \bibinfo{year}{2013}.
\newblock \bibinfo{title}{Teichm{\"u}ller shape descriptor and its application
  to alzheimer's disease study}.
\newblock \bibinfo{journal}{Int. J. Comput. Vision} \bibinfo{volume}{105},
  \bibinfo{pages}{155--170}.
\bibitem[{Zhang et~al.(2017a)Zhang, Fan, Li, Thompson, Ye and
  Wang}]{Zhang:ISBI17}
\bibinfo{author}{Zhang, J.}, \bibinfo{author}{Fan, Y.}, \bibinfo{author}{Li,
  Q.}, \bibinfo{author}{Thompson, P.M.}, \bibinfo{author}{Ye, J.},
  \bibinfo{author}{Wang, Y.}, \bibinfo{year}{2017}a.
\newblock \bibinfo{title}{Empowering cortical thickness measures in clinical
  diagnosis of {A}lzheimer's disease with spherical sparse coding}.
\newblock \bibinfo{journal}{Proc IEEE Int Symp Biomed Imaging}
  \bibinfo{volume}{2017}, \bibinfo{pages}{446--450}.
\bibitem[{Zhang et~al.(2017b)Zhang, Fan, Li, Thompson, Ye and
  Wang}]{zhang2017empowering}
\bibinfo{author}{Zhang, J.}, \bibinfo{author}{Fan, Y.}, \bibinfo{author}{Li,
  Q.}, \bibinfo{author}{Thompson, P.M.}, \bibinfo{author}{Ye, J.},
  \bibinfo{author}{Wang, Y.}, \bibinfo{year}{2017}b.
\newblock \bibinfo{title}{Empowering cortical thickness measures in clinical
  diagnosis of alzheimer's disease with spherical sparse coding}, in:
  \bibinfo{booktitle}{2017 IEEE 14th International Symposium on Biomedical
  Imaging (ISBI 2017)}, \bibinfo{organization}{IEEE}. pp.
  \bibinfo{pages}{446--450}.
\bibitem[{Zhang et~al.(2017c)Zhang, Li, Caselli, Thompson, Ye and
  Wang}]{zhang2017multi}
\bibinfo{author}{Zhang, J.}, \bibinfo{author}{Li, Q.},
  \bibinfo{author}{Caselli, R.J.}, \bibinfo{author}{Thompson, P.M.},
  \bibinfo{author}{Ye, J.}, \bibinfo{author}{Wang, Y.}, \bibinfo{year}{2017}c.
\newblock \bibinfo{title}{{{M}ulti-Source Multi-Target Dictionary Learning for
  Prediction of Cognitive Decline}}.
\newblock \bibinfo{journal}{Inf Process Med Imaging} \bibinfo{volume}{10265},
  \bibinfo{pages}{184--197}.
\bibitem[{Zhang et~al.(2016a)Zhang, Shi, Stonnington, Li, Gutman, Chen, Reiman,
  Caselli, Thompson, Ye et~al.}]{zhang2016hyperbolic}
\bibinfo{author}{Zhang, J.}, \bibinfo{author}{Shi, J.},
  \bibinfo{author}{Stonnington, C.}, \bibinfo{author}{Li, Q.},
  \bibinfo{author}{Gutman, B.A.}, \bibinfo{author}{Chen, K.},
  \bibinfo{author}{Reiman, E.M.}, \bibinfo{author}{Caselli, R.},
  \bibinfo{author}{Thompson, P.M.}, \bibinfo{author}{Ye, J.}, et~al.,
  \bibinfo{year}{2016}a.
\newblock \bibinfo{title}{Hyperbolic space sparse coding with its application
  on prediction of {A}lzheimer’s disease in {M}ild {C}ognitive {I}mpairment},
  in: \bibinfo{booktitle}{Med Image Comput Comput Assist Interv},
  \bibinfo{organization}{Springer}. pp. \bibinfo{pages}{326--334}.
\bibitem[{Zhang et~al.(2016b)Zhang, Stonnington, Li, Shi, Bauer, Gutman, Chen,
  Reiman, Thompson, Ye and Wang}]{Zhang:ISBI2016}
\bibinfo{author}{Zhang, J.}, \bibinfo{author}{Stonnington, C.},
  \bibinfo{author}{Li, Q.}, \bibinfo{author}{Shi, J.}, \bibinfo{author}{Bauer,
  R.J.}, \bibinfo{author}{Gutman, B.A.}, \bibinfo{author}{Chen, K.},
  \bibinfo{author}{Reiman, E.M.}, \bibinfo{author}{Thompson, P.M.},
  \bibinfo{author}{Ye, J.}, \bibinfo{author}{Wang, Y.}, \bibinfo{year}{2016}b.
\newblock \bibinfo{title}{Applying sparse coding to surface multivariate
  tensor-based morphometry to predict future cognitive decline}.
\newblock \bibinfo{journal}{Proc IEEE Int Symp Biomed Imaging}
  \bibinfo{volume}{2016}, \bibinfo{pages}{646--650}.
\bibitem[{Zhang et~al.(2018a)Zhang, Tu, Li, Caselli, Thompson, Ye and
  Wang}]{zhang2018multi}
\bibinfo{author}{Zhang, J.}, \bibinfo{author}{Tu, Y.}, \bibinfo{author}{Li,
  Q.}, \bibinfo{author}{Caselli, R.J.}, \bibinfo{author}{Thompson, P.M.},
  \bibinfo{author}{Ye, J.}, \bibinfo{author}{Wang, Y.}, \bibinfo{year}{2018}a.
\newblock \bibinfo{title}{Multi-task sparse screening for predicting future
  clinical scores using longitudinal cortical thickness measures}, in:
  \bibinfo{booktitle}{Proc IEEE Int Symp Biomed Imaging}, pp.
  \bibinfo{pages}{1406--1410}.
\bibitem[{Zhang et~al.(2018b)Zhang, Lv, Li, Zhu, Jiang, Zhang, Zhao, Guo, Ye,
  Hu and Liu}]{pmid29993466}
\bibinfo{author}{Zhang, W.}, \bibinfo{author}{Lv, J.}, \bibinfo{author}{Li,
  X.}, \bibinfo{author}{Zhu, D.}, \bibinfo{author}{Jiang, X.},
  \bibinfo{author}{Zhang, S.}, \bibinfo{author}{Zhao, Y.},
  \bibinfo{author}{Guo, L.}, \bibinfo{author}{Ye, J.}, \bibinfo{author}{Hu,
  D.}, \bibinfo{author}{Liu, T.}, \bibinfo{year}{2018}b.
\newblock \bibinfo{title}{Experimental comparisons of sparse dictionary
  learning and independent component analysis for brain network inference from
  f{M}{R}{I} data}.
\newblock \bibinfo{journal}{IEEE Trans Biomed Eng} .

\end{thebibliography}

\end{document}